\documentclass[runningheads,orivec]{llncs}
\usepackage[T1]{fontenc}
\usepackage{hyperref}
\usepackage{amsmath,amssymb,amsfonts}
\usepackage{stmaryrd}
\usepackage{mathrsfs}
\usepackage{todonotes}

\usepackage{tikz}
\usetikzlibrary{automata}
\usetikzlibrary{positioning}
\usetikzlibrary{fit}
\usetikzlibrary{calc}
\usetikzlibrary{backgrounds}
\usetikzlibrary{shapes}
\usetikzlibrary{arrows}
\usepackage{array,multirow,graphicx}
\usepackage{comment}

\newcommand{\logikey}{\textsc{LogiKEy}}

\begin{document}

\title{Faithful Logic Embeddings in HOL --- Deep and Shallow}

\author{Christoph Benzm\"uller\inst{1,2}\orcidID{0000-0002-3392-3093} }

\institute{Otto-Friedrich-Universit\"at Bamberg, Kapuzinerstraße 16, 96047 Bamberg, Germany \and
Freie Universit\"at Berlin, Kaiserswerther Str. 16-18
14195 Berlin, Germany \\
\email{christoph.benzmueller@uni-bamberg.de} $\mid$ \email{c.benzmueller@fu-berlin.de}
}

\maketitle              

\begin{abstract}
Deep and shallow embeddings of non-classical logics in classical higher-order logic have been explored, implemented, and used in various reasoning tools in recent years. This paper presents a method for the simultaneous deployment of deep and shallow embeddings of various degrees in classical higher-order logic. This enables flexible, interactive and automated theorem proving and counterexample finding at meta and object level, as well as automated faithfulness proofs between these logic embeddings. The method is beneficial for logic education, research and application and is illustrated here using a simple propositional modal logic. However, this approach is conceptual in nature and not limited to this simple logic context.


%\keywords{Deep and shallow embeddings \and Non-classical logics \and Higher-order logic \and Faithfulness \and Automated Reasoning \and Proof assistants}
\end{abstract}

\section{Motivation and Introduction} 
\label{sec:introduction} 
Deep embeddings of logics, or more generally of domain-specific languages, in a suitable metalogic, such as classical higher-order logic (HOL) \cite{church1940,B28}, are typically based on explicitly introduced abstract data types. They essentially axiomatically characterize the structure and the meaning of the new language.  Shallow embeddings, on the other hand, identify the embedded languages directly as fragments of the metalogic and they typically require few or no axiom postulates.\footnote{Gibbons and Wu characterize the two notions as follows \cite[p.2]{GibbonsW14}: \emph{``[...] a deep embedding consists of a representation of the abstract syntax as an algebraic datatype, together with some functions that assign semantics to that syntax by traversing the algebraic datatype. A shallow embedding eschews the algebraic datatype, and hence the explicit representation of the abstract syntax of the language; instead, the language is defined directly in terms of its semantics.''}} Deep and shallow embeddings of logics in HOL effectively and elegantly support interactive and automated reasoning with and about such logics within existing HOL theorem proving systems. Consequently, they have received increased attention in recent years. Shallow embeddings, for example, are well suited to automate even quantified modal logics, enabling applications in areas such as computational metaphysics \cite{J75}. Deep embeddings, on the other hand, are well suited to enable reasoning about the syntactic structure of embedded logics.

The contributions of this paper are manifold: First, a technique for the simultaneous provision of deep and shallow embeddings in metalogic HOL is presented, with the effect that computer-supported faithfulness proofs between the embeddings are not only enabled, but may --- at least for simple embedded logics --- now even be fully automated in HOL. This is, to the best of the authors' knowledge, a novel result. Second, this paper discusses shallow embedding alternatives differing in the degree of semantic dependencies that are explicitly encoded and maintained. More precisely, heavyweight (maximal) and lightweight (minimal) shallow embeddings are contrasted as extremes, and doing so, the paper hints at the full spectrum that exists between these extremes, which is in fact relevant for practice. For example, while a minimal shallow embedding for a normal modal logic only needs to explicitly encode the crucial semantic dependency on possible worlds~\cite{J21}, the explicit maintenance of an additional dependency, namely on dynamically updated domains of possible worlds, becomes relevant in case the logic is extended by a public announcement operator \cite{J58}. Third, the paper hints at opportunities for logic education, and for research, about and with logic embeddings within modern proof assistants and theorem provers for HOL. These range from metalogical studies to object logic applications using interactive proof and disproof, as well as automated theorem proving and counterexample finding, while allowing precise control over the particular level (object- or metalevel) at which  reasoning takes place. Fourth, the paper presents some preliminary studies on whether shallow or deep embeddings are better suited for certain reasoning tasks. Fifth, using the Löb axiom of provability logic, the paper briefly sketches how research on the detection of semantic correspondences could possibly be supported by the presented technique in the future.

Since this paper and associated material is intended to be used in logic education, simple propositional modal logic (PML) \cite{sep-logic-modal,BlackburnB07,Sider2009} has been chosen to illustrate the addressed ideas, and Isabelle/HOL \cite{Isabelle} has been chosen as the host proof assistant system for the experiments conducted. Reasons for using Isabelle/HOL include (i) its very good proof automation support with e.g.~the \texttt{sledgehammer} \cite{sledgehammer} tool, (ii) the availability of the (counter-)model finder \texttt{nitpick} \cite{nitpick}, and (iii) its powerful user interface, which allows for user-friendly \LaTeX-quality representations. In principle, this paper could have taken advantage of Isabelle/HOL's special support for computer-verifiable paper writing to generate a document directly from the Isabelle/HOL sources, as is the case, for example, with all the articles published in the Archive of Formal Proofs. However, the author has decided against this idea and in favor of screenshots of the source code in order to save space. Furthermore, the visual highlighting and indenting in these screenshots makes them more useful for teaching and illustration purposes by drawing attention to the most important information.

Organization of the paper: \S\ref{sec:pml} and \S\ref{sec:hol} briefly introduce PML and HOL, before \S\ref{sec:pmlinhol_deep}-\ref{sec:pmlinhol_shallow_min} present deep, maximally shallow, and minimally shallow embeddings of PML in HOL. These are connected by automated faithfulness proofs given in \S\ref{sec:pmlinhol_faithfulness}. Various experiments with the presented embeddings are presented in \S\ref{sec:pmlinhol_experiments}, before \S\ref{sec:logikey} discusses related work and \S\ref{sec:conclusion} concludes the paper.

\section{Propositional modal logic (PML)}
\label{sec:pml}
The \emph{syntax} of PML is defined by the grammar $\varphi, \psi := a \mid \neg \varphi  \mid  \varphi \supset \psi  \mid \Box \varphi$, where $a\in\mathcal{S}$ are propositional symbols declared in a (nonempty) signature $\mathcal{S}$. Further logical connectives can be defined as usual, e.g.: $\varphi \vee \psi := \neg \varphi \supset \psi$, $\varphi \wedge \psi := \neg (\varphi \supset \neg \psi)$, $\Diamond \varphi := \neg \Box \neg \varphi$, $\top := p \supset p$ (for some $p\in\mathcal{S}$), and $\bot :=\neg \top$.

A \emph{semantics} for PML, called relational semantics or possible-world semantics, determines the truth of a formula $\varphi$ relative to a so-called possible world $w$ and a relational model $\mathcal{M}$. A relational model $\mathcal{M}$ is a tuple $\langle \mathcal{W},\mathcal{R},\mathcal{V}\rangle$, where
$\mathcal{W}$ is a set of possible worlds (the universe), 
$\mathcal{R}$ is a binary relation on $W$ (called the accessibility relation),
$\mathcal{V}$ is a valuation function that assigns a truth value to each symbol $a \in \mathcal{S}$ in each world $w\in \mathcal{W}$, where $\mathcal{S}$ is the signature of propositional symbols. $\langle \mathcal{W},\mathcal{R}\rangle$ is called a \emph{frame}.
Formally, the truth of a formula $\varphi$ with respect to a given model $\langle \mathcal{W},\mathcal{R},\mathcal{V}\rangle$ and a possible world $w\in \mathcal{W}$ is defined as follows:
\begin{align*}
    \langle \mathcal{W},\mathcal{R},\mathcal{V}\rangle, w & \models a \text{\ iff\ } \mathcal{V}(a)(w) \text{\ is true} \\
    \langle \mathcal{W},\mathcal{R},\mathcal{V}\rangle, w & \models \neg \varphi \text{\ iff\ not\ } \langle \mathcal{W},\mathcal{R},\mathcal{V}\rangle, w \models \varphi \\
    \langle \mathcal{W},\mathcal{R},\mathcal{V}\rangle, w & \models \varphi \supset \psi \text{\ iff\ } \langle \mathcal{W},\mathcal{R},\mathcal{V}\rangle, w \models \varphi \text{\ implies\ }\langle \mathcal{W},\mathcal{R},\mathcal{V}\rangle, w \models \psi \\
    \langle \mathcal{W},\mathcal{R},\mathcal{V}\rangle, w & \models \Box \varphi \text{\ iff\ for\ all\ } v \text{\ with\ } \mathcal{R} w v \text{\ we have\ }  \langle \mathcal{W},\mathcal{R},\mathcal{V}\rangle, v \models \varphi 
\end{align*}

Note that in this recursive definition of truth the only parameter altered is the possible world $w$ in the $\Box$-clause, which states that a formula $\varphi$ is `necessary' in a world $w$ if and only if (iff) $\varphi$ is true at all worlds $v$ that are reachable from $w$ via the accessibility relation $\mathcal{R}$. 
Based on the above definitions a notion of validity can be defined:
A formula $\varphi$ is valid (in so called base modal logic \textbf{K}) iff $\langle \mathcal{W},\mathcal{R},\mathcal{V}\rangle, w \models \varphi$ for all $\mathcal{W}$, $\mathcal{R}$, $\mathcal{V}$, and $w\in\mathcal{W}$.

Different systems of modal logic beyond base logic \textbf{K} can be obtained by requiring (combinations) of so called \emph{frame conditions}. For example, a frame is called (i) reflexive iff $Rxx$ for all $x\in \mathcal{W}$, (ii) symmetric iff  $Rxy$ implies $Ryx$, for all for all $x,y\in \mathcal{W}$, and (iii) transitive iff $Rxy$ and $Ryz$ implies $Rxz$, for all for all $x,y,z\in \mathcal{W}$. Moreover, there are well-known correspondences, known as Sahlqvist correspondences~\cite{Sahlqvist1975},  between such semantic conditions and the validity of certain axiom schemata. For example, (i) corresponds to the validity of axiom schema M: $\Box \varphi \supset \varphi$, (ii) to B: $\varphi \supset \Box \Diamond \varphi$ and (iii) to 4: $\Box \varphi \supset \Box \Box \varphi$.

A \emph{proof system} for modal logic \textbf{K} is obtained by adding the schematic necessitation rule, stating that `if $\varphi$ is a theorem of \textbf{K}, then so is $\Box \varphi$' and the distribution axiom schema $\Box (\varphi \supset \psi) \supset (\Box \varphi \supset \Box \psi)$ to the principles of classical propositional logic.
The latter can be captured by a simple Hilbert calculus \cite{HilbertSystem,smullyan2014beginner} consisting of the modes ponens rule, `if $\varphi$ and $\varphi\supset\psi$ are theorems (of \textbf{K}), then so is $\psi$', and the three axiom schemata H1: $\varphi \supset (\psi \supset \varphi)$, H2: $(\varphi \supset (\psi \supset \gamma)) \supset ((\varphi \supset \psi)\supset (\varphi \supset \gamma))$ and H3: $(\neg \varphi \supset \neg \psi) \supset (\psi \supset \varphi)$).

Well known modal logics beyond \textbf{K} (which works without additional frame conditions) include the logic \textbf{S4} with e.g.~the frame conditions (i) and (iii) and the logic \textbf{S5} with e.g.~the frame conditions (i), (ii) and (iii).
Thus, by adding the schematic axioms (M)+(4) or (M)+(B)+(4) to the proof system for modal logic \textbf{K} above, proof systems for \textbf{S4} and \textbf{S5} are obtained.

\section{Classical higher-order logic (HOL)}
\label{sec:hol}
The \emph{syntax} of HOL is defined by
$s,t := p_\alpha \ | \ X_\alpha \ | \ (\lambda X_\alpha s_\beta)_{\alpha {\Rightarrow} \beta} \ | \ (s_{\alpha{\Rightarrow}\beta}t_\alpha)_\beta$,
where $\alpha, \beta, o \in \mathcal{T}$ and with $\mathcal{T}$ being a set of \emph{simple types} defined by $\alpha , \beta := o \ | \ i \ | \ (\alpha{\Rightarrow}\beta)$. Type $o$, called $\texttt{bool}$ in Isabelle/HOL, denotes truth values, $i$ individuals, and ${\Rightarrow}$ is the function type constructor.
The $p_\alpha$ in the syntax of HOL are typed constants symbols declared in the sets $S_\alpha$ of a given signature $\mathscr{S}=(S_\alpha)_{\alpha\in\mathcal{T}}$. The $X_\alpha$  are typed variables (distinct from the $p_\alpha$). 
Complex HOL terms are constructed from given HOL terms via $\lambda$-abstraction $(\lambda X_\alpha s_\beta)_{\alpha {\Rightarrow} \beta}$ and function application $(s_{\alpha {\Rightarrow} \beta}t_\alpha)_\beta$. %
HOL is thus a logic of terms defined on top of the simply typed $\lambda$-calculus, and terms of type $o$ are called \emph{formulas}. The type of each term is given as a subscript and may be omitted if obvious in context. 
It is assumed that signature $\mathscr{S}$ contains the following 
\emph{primitive logical connectives}: $\neg_{o {\Rightarrow} o}, \vee_{o{\Rightarrow} o{\Rightarrow} o}$, $=_{\alpha {\Rightarrow} \alpha {\Rightarrow} \alpha }$ (short: $=^\alpha$) and $\Pi_{(\alpha {\Rightarrow} o) {\Rightarrow} o}$ (short: $\Pi^\alpha$).
Further logical connectives or shorthand notations can be introduced as usual; e.g.~$\longrightarrow_{o{\Rightarrow} o{\Rightarrow} o} \, := \lambda X_o \lambda Y_o (\neg X \vee Y$) and $\forall X_\alpha\, \varphi_o := \Pi^\alpha (\lambda X_\alpha\, \varphi)$.

The definition of a \emph{semantics} for HOL starts with the notion of 
a \emph{frame} $\mathcal{D}$, a collection $\{\mathcal{D}_\alpha\}_{\alpha \in T}$ of nonempty sets $\mathcal{D}_\alpha$, such that $\mathcal{D}_o = \{T, F\}$ (for the truth values true and false). 
$\mathcal{D}_i$ is chosen freely and $\mathcal{D}_{\alpha {\Rightarrow} \beta}$ are collections of functions mapping $\mathcal{D}_\alpha$ into $\mathcal{D}_\beta$.
A \emph{model} for HOL is a tuple $\mathcal{M} = \langle \mathcal{D}, I \rangle$, where $\mathcal{D}$ is a frame, and $I$ is a family of typed interpretation functions mapping constant symbols $p_\alpha \in C_\alpha$ to appropriate elements of $\mathcal{D}_\alpha$, called the \emph{denotation} of $p_\alpha$.
The logical connectives $\neg, \vee, \Pi$ and $=$ are given their expected standard denotations, for example, $I(\neg)$ is such that $I(\neg)(T) = F$ and  $I(\neg)(F) = T$, and $I(\Pi^\alpha)$ is such that for all $s \in \mathcal{D}_{\alpha {\Rightarrow} o}$ we have $ I(\Pi^\alpha)(s) = T$ iff $s(a) = T$ for all $a \in \mathcal{D}_\alpha$.

A \emph{variable assignment g} maps variables $X_\alpha$ to elements in $\mathcal{D}_\alpha$, and $g[d/W]$ denotes the $g'$ identical to $g$, except for variable $W$, which is now mapped to $d$.
	
The \emph{denotation} $\llbracket s_\alpha\rrbracket^{M,g}$ of an HOL term $s_\alpha$ on a model $\mathcal{M} = \langle \mathcal{D}, I \rangle$ under assignment $g$ is an element $d \in \mathcal{D}_\alpha$ defined by $\llbracket p_\alpha\rrbracket^{\mathcal{M}, g}$ = $I(p_\alpha)$, $\llbracket X_\alpha\rrbracket^{\mathcal{M}, g}$ = $g(X_\alpha)$, $\llbracket (s_{\alpha {\Rightarrow} \beta}t_\alpha)_\beta\rrbracket^{\mathcal{M}, g}$  = $\llbracket s_{\alpha {\Rightarrow} \beta}\rrbracket^{\mathcal{M}, g}(\llbracket t_\alpha\rrbracket^{\mathcal{M}, g})$, and $\llbracket (\lambda X_\alpha s_\beta)_{\alpha {\Rightarrow} \beta}\rrbracket^{\mathcal{M}, g}$  =  the $f:\mathcal{D}_\alpha \longrightarrow \mathcal{D}_\beta$ s.t. $\forall d \in \mathcal{D}_\alpha: f(d) = \llbracket s_\beta \rrbracket^{\mathcal{M},g[d/X_\alpha]}$. 
Note that from the given definitions it e.g.~follows $\llbracket \forall X_\alpha \varphi_o\rrbracket^{\mathcal{M}, g}  = T \text{ iff } \llbracket  \varphi_o\rrbracket^{\mathcal{M}, g[d/X_\alpha]} = T \text{ for all } d\in \mathcal{D}_\alpha$.

In a \emph{standard model} a domain $\mathcal{D}_{\alpha {\Rightarrow} \beta}$ is defined as the set of all total functions from $\mathcal{D}_\alpha$ to $\mathcal{D}_\beta$:
$\mathcal{D}_{\alpha {\Rightarrow} \beta} = \{ f \ |\ f : \mathcal{D}_\alpha {\Rightarrow} \mathcal{D}_\beta \}$. 
In a \emph{Henkin model} (or general model) \cite{henkin1950,Andrews72a} function spaces are not necessarily required to be the full set of functions: $\mathcal{D}_{\alpha {\Rightarrow} \beta} \subseteq \{ f \ |\ f : \mathcal{D}_\alpha {\Rightarrow} \mathcal{D}_\beta \}$.  However, it is required 
that every term still denotes.

$s_o$ is \emph{valid in $\mathcal{M}$ under assignment $g$}, denoted as $\mathcal{M},g \models^\texttt{HOL} s_o$,  iff $\llbracket s_o\rrbracket^{\mathcal{M},g}${$=T$}
$s_o$ is \emph{valid in $\mathcal{M}$}, denoted as $\mathcal{M} \models^\texttt{HOL} s_o$, iff $\mathcal{M},g \models^\texttt{HOL}s_o$ for all assignments $g$, and 
$s_o$ is \emph{valid}, denoted as $\models^\texttt{HOL}s_o$,  iff $s_o$ is valid in all Henkin models $\mathcal{M}$.  

Each standard model is obviously also a Henkin model. Consequently, if $\models^\texttt{HOL}s_o$, then $s_o$ is also valid in all standard models.

Due to G\"odel \cite{goedel1931}, sound and complete mechanizations of HOL with standard semantics cannot be achieved. For HOL with Henkin semantics, however, sound and complete calculi exist; see e.g.~\cite{J6,B5}. Relevant for this paper is that Isabelle/HOL is also sound and complete for HOL, and that HOL as presented is sufficiently well suited as a metalogic for the work presented in this paper.

\section{Deep embedding of PML in HOL}
\label{sec:pmlinhol_deep}
\begin{figure}[tp!]
  \centering
  \colorbox{gray!30}{\includegraphics[width=.97\textwidth]{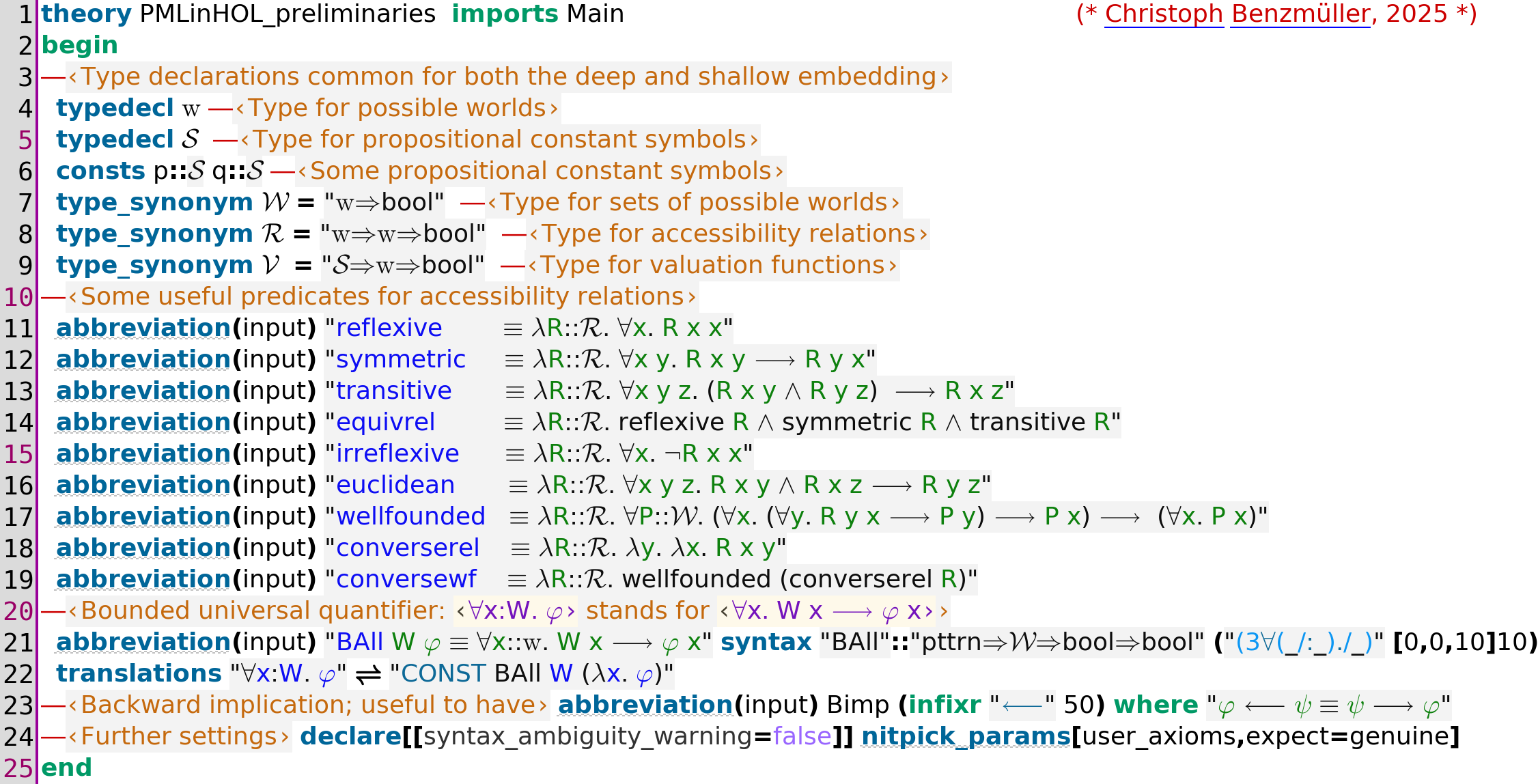}}
  \caption{Preliminaries.}
  \label{fig:Preliminaries}
\end{figure}

This section presents a deep embedding of PML in HOL. Both the PML language and its semantics are formalized very closely following the definitions given in~\S\ref{sec:pml}. This formalization, and all others presented in this paper, are based on some uniformly assumed conventions, resp.~preliminaries. These preliminaries are provided in a file \texttt{PMLinHOL\_preliminaries.thy}, see Fig~\ref{fig:Preliminaries}. Useful comments are provided; they are encapsulated in orange delimiters \textcolor{orange}{${-}{<}\ {>}$}.

Relevant at this point is that a type $\mathcal{S}$ is introduced to represent the signature $\mathcal{S}$ of PML, i.e. a (finite or infinite) set of propositional logic symbols; type $\texttt{w}$ is the type of possible worlds, $\mathcal{W} : = \texttt{w} {\Rightarrow} \texttt{bool}$ is the predicate type of possible worlds (we regard predicates of this type as characteristic functions for the associated sets of possible worlds), $\mathcal{R} := \texttt{w} {\Rightarrow} \texttt{w} {\Rightarrow} \texttt{bool}$ is the type of accessibility relations between possible worlds, and $\mathcal{V} := \mathcal{S} {\Rightarrow} \texttt{w} {\Rightarrow} \texttt{bool}$ is the type of evaluation functions
(these definitions are shown in lines 4-9). Well known predicates and functions on relations are introduced (in lines 11-19), and 
$\forall w{:}W.\, \varphi$ (see lines 21-22) is a $W$-predicate guarded quantifier that unfolds into $\forall w. \,W\,w \longrightarrow \varphi$, where $W$ is of type $\mathcal{W}$. 
 An important remark concerns the sharing of, e.g., the signature type $\mathcal{S}$ between all embeddings introduced in this paper. This sharing is essential for enabling the faithfulness proofs between the embeddings as presented in \S\ref{sec:pmlinhol_faithfulness}.

\begin{figure}[tp!]
  \centering
  \colorbox{gray!30}{\includegraphics[width=.97\textwidth]{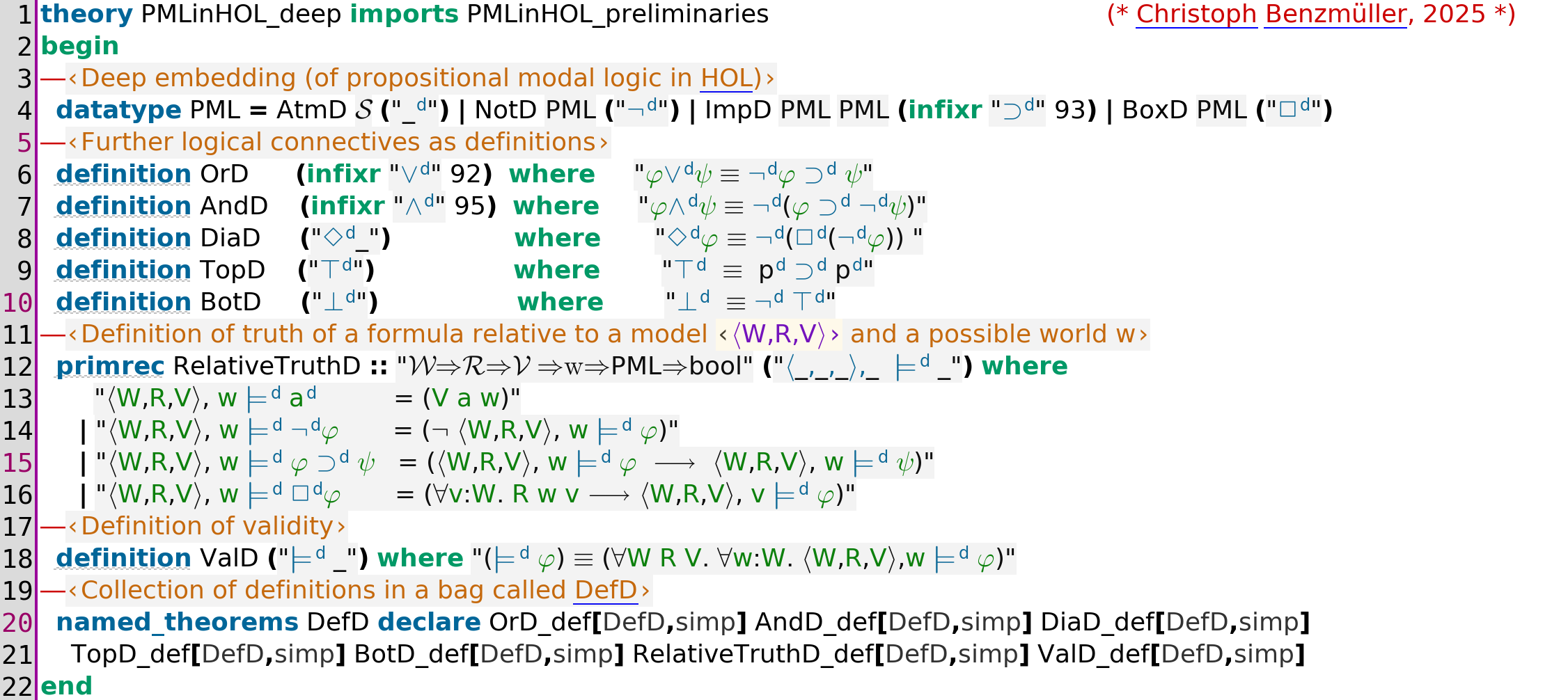}}
    \vskip1em

  \colorbox{gray!30}{\includegraphics[width=.97\textwidth]{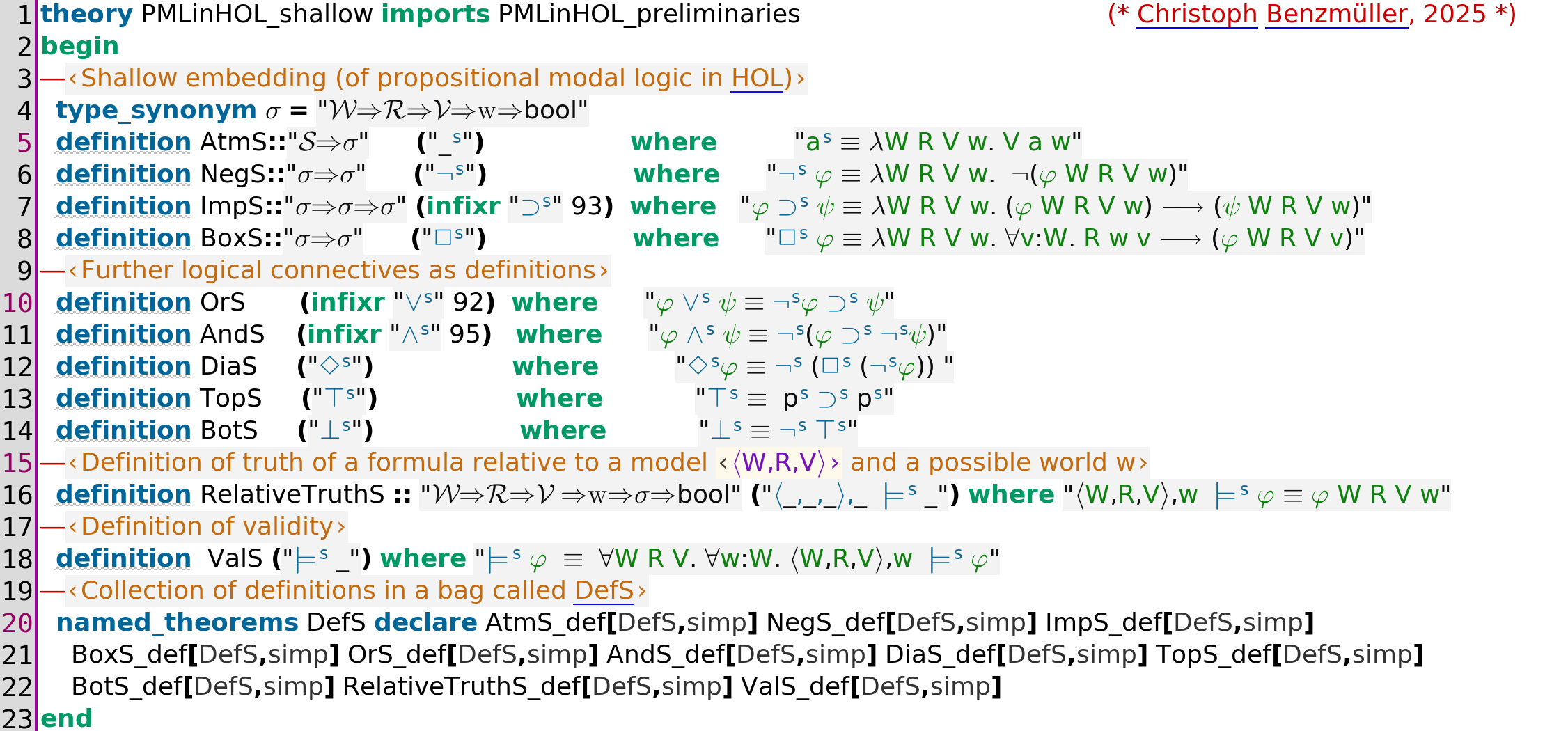}}
    \vskip1em

  \colorbox{gray!30}{\includegraphics[width=.97\textwidth]{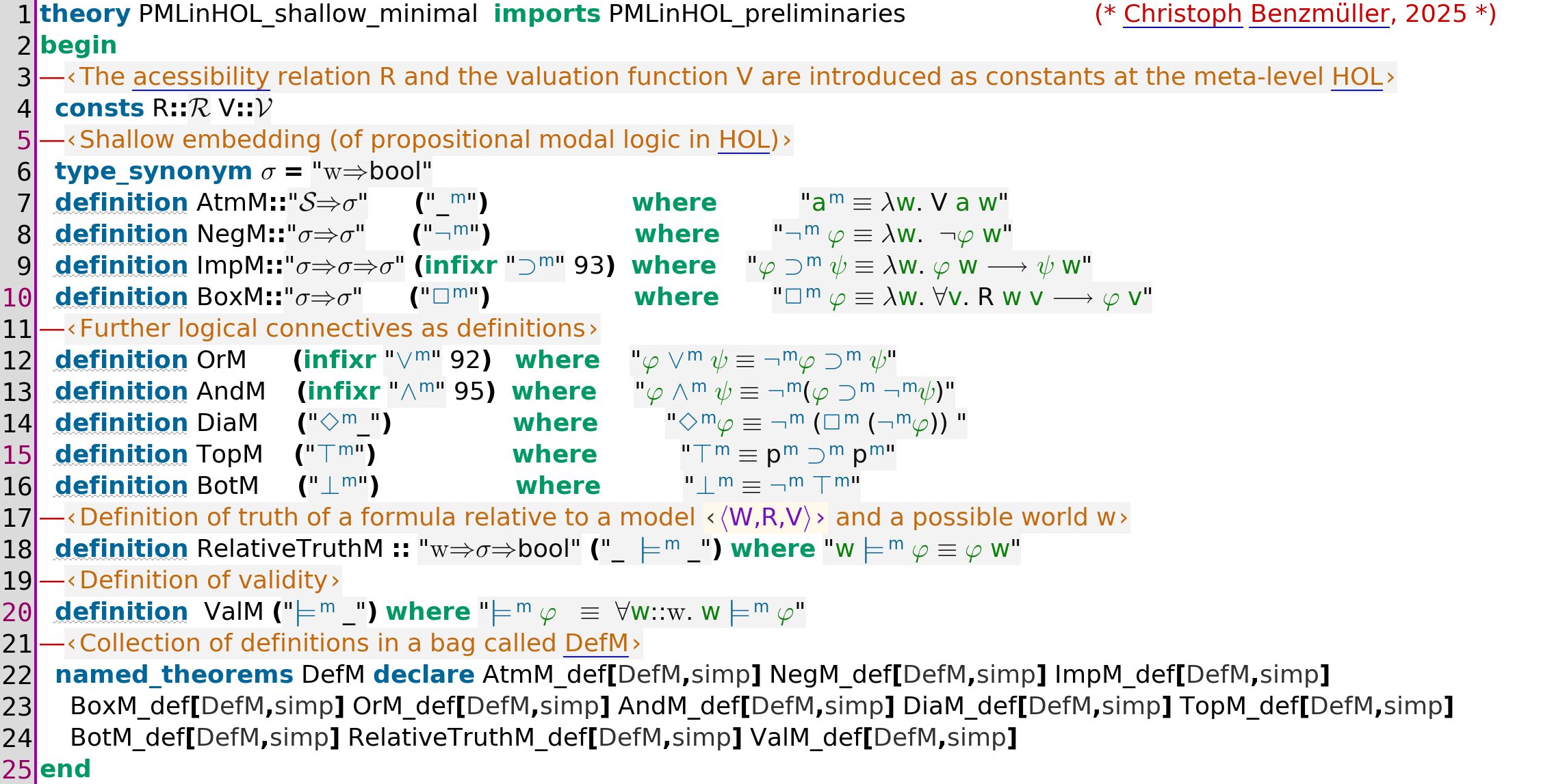}}
  \caption{PML in HOL: deep (top), maximal shallow (mid), and minimal shallow (bot).\label{fig:PMLinHOL_all}}
\end{figure}

The deep embedding of PML in HOL is defined in file  \texttt{PMLinHOL\_deep.thy} and presented at top of Fig.~\ref{fig:PMLinHOL_all}. After importing \texttt{PMLinHOL\_preliminaries.thy}, this file starts out with defining an abstract datatype\footnote{Such a datatype declaration corresponds to, and unfolds into, a rather large set of axioms in HOL that could be postulated instead. Using the Isabelle/HOL command \texttt{print\_theorems}, this set can be inspected by the user.} introducing the inductively defined logic language PML, which introduces atomic formulas $a^d$ (for $a\in\mathcal{S}$), negated formulas $\neg^d \varphi$, implication formulas $\varphi \supset^d \psi$, and boxed formulas $\Box^d \varphi$ (see line 4). The superscript $^d$ is used to mark all deeply embedded connectives so that they can be properly distinguished from their counterparts defined in the following sections. In these sections, the alternative superscripts~$^s$ (maximal shallow) and $^m$ (minimal shallow) are used accordingly.

In accordance with \S\ref{sec:pml} the
 further logical connectives $\vee^d$, $\wedge^d$, $\Diamond^d$, $\top^d$, and $\bot^d$ are introduced as defined terms (lines 6-10).\footnote{Abbreviations could be used instead. They are always unfolded implicitly, while definition unfolding is an explicit operation, leading to more options for control. In any case, readers new to Isabelle/HOL can simply focus on the indented definition equations here (and in other figures), at least at first reading.}  Then the primitive recursive definition of truth in a world $w$ for a given model $\langle {W}, {R}, {V} \rangle$, termed \texttt{RelativeTruthD} and noted as $\langle {W}, {R}, {V} \rangle, w \models^d \varphi$, is given (lines 12-16). This definition follows one to one, even in presentation style, the definition introduced in \S\ref{sec:pml}. Based on this, the validity $\models^d \varphi$ of a deeply embedded formula $\varphi$ is defined as relative truth w.r.t.~all sets of worlds $W$, accessibility relations $R$, valuation functions $V$, and worlds $w\in W$ (line 18). Finally, in order to support a convenient, simultaneous unfolding of the introduced definitions, a bag \texttt{DefD} is defined (in lines 20-21), which bundles them and adds them also to the simplifier \texttt{simp} of Isabelle/HOL.

\section{Maximal shallow embedding of PML in HOL}
\label{sec:pmlinhol_shallow_max}

The shallow embedding of PML in HOL presented in the middle part of Fig.~\ref{fig:PMLinHOL_all}, which is
formalized in file \texttt{PMLinHOL\_shallow.thy}, is a \emph{maximal} (heavyweight) shallow embedding in the following sense: PML formulas $\varphi$ are associated here with HOL predicate terms $\varphi_\sigma$ of type $\sigma := \mathcal{W} {\Rightarrow} \mathcal{R} {\Rightarrow} \mathcal{V} {\Rightarrow} \texttt{w} {\Rightarrow} \texttt{bool}$ (see line 4), thus explicitly encoding the entire semantic dependencies of a PML formula $\varphi_\sigma$ on a set of worlds ${W}$, an accessibility relation ${R}$, a valuation function ${V}$ and a world $w$. Remember, that $\mathcal{W}$, $\mathcal{R}$, $\mathcal{V}$, and $\texttt{w}$ have been uniformly declared for all embeddings presented in this paper in the 
imported preliminaries file.

Exploiting $\lambda$-abstraction and the logical connectives $\neg$ and $\longrightarrow$ from metalogic HOL, the definitions of deeply embedded negation $\neg^s$ and implication $\supset^s$ (in lines 6-8), recursively pass on all the dependencies as mentioned:\footnote{Dot-notation is used, whereby the scope 
opened by $.$ is reaching as far to the right as consistent with the formula structure. The term $\lambda W R\,V\,w.\, \neg (\varphi\, W R\,V\,w)$ thus corresponds to $\lambda W \lambda R\lambda V \lambda w (\neg (\varphi\, W R\,V\,w))$.} $\neg^s \varphi := \lambda W R\,V\,w.\, \neg (\varphi\, W R\,V\,w)$ and $\varphi \supset^s \psi := \lambda W R\,V\,w.\, (\varphi\, W R\,V\,w) \longrightarrow (\psi\, W R\,V\,w)$. Atoms $a^s$, for $a\in \mathcal{S}$, are evaluated (in line 5) using the explicitly maintained valuation  function $V$, that is, $a^s := \lambda W R\,V\,w.\, V\,a\,w$.
The only case where parts of the recursively passed on dependencies are actually modified is (in line 8) for $\Box^s \varphi$, since here the world in which $\varphi$ is recursively evaluated may change with the transition, via the accessibility relation $R$, to world $v$: $\Box^s \varphi := \lambda W R\,V\,w.\, \forall v. R\,w\,v \longrightarrow (\varphi\, W R\,V\,v)$. The definition of the other logical connectives $\vee^s$, $\wedge^s$, $\Diamond^s$, $\top^s$ and $\bot^s$ (in lines 10-14) is analogous to the deep embedding and as given in \S\ref{sec:hol}. 

Truth in a world $w$ for a given model $\langle {W}, {R}, {V} \rangle$, called \texttt{RelativeTruthS} and noted as $\langle {W}, {R}, {V} \rangle, w \models^s \varphi$, is then defined (see line 16) as the application of $\varphi$ to ${W}$, ${R}$, ${V}$ and $w$. Analogous to the deep embedding from above, validity $\models^s \varphi$ of an embedded formula $\varphi$ is defined as relative truth w.r.t.~all sets of worlds $W$, accessibility relations $R$, valuation functions $V$, and worlds $w\in W$ (line 18). 
Also analogous to above, a bag \texttt{DefS} is defined (lines 20-22), which conveniently bundles all the definitions given in this file and adds them to the simplifier. Note that not a single axiom is being introduced in this maximal shallow embedding, with the effect, that PML is being characterized here as a proper, albeit quite heavyeight, fragment of the language of HOL.

\section{Minimal shallow embedding of PML in HOL}
\label{sec:pmlinhol_shallow_min}

The maximal shallow embedding presented in \S\ref{sec:pmlinhol_shallow_max} explicitly maintains more dependencies than seemingly necessary,  since the set of worlds ${W}$, the accessibility relation ${R}$, and the valuation function ${V}$ are never modified in any of the given definitions, but simply passed on. Therefore, it makes sense to treat these static dependencies as parameters in the metalogic HOL. This is the core idea of the \emph{minimal} (leightweight) shallow embedding of PML in HOL as shown at the bottom of Fig.~\ref{fig:PMLinHOL_all} and formalized in file \texttt{PMLinHOL\_shallow\_minimal.thy}. 

To implement this idea, ${W}$ is now implicitly identified with type $\texttt{w}$ of HOL at the metalevel, and the constant symbols $\texttt{R}$ and $\texttt{V}$ are introduced in metalogic HOL to represent the semantic notions ${R}$ and ${V}$ of PML (see line 4). Most importantly, the modified HOL type $\sigma := \texttt{w}{\Rightarrow}\texttt{bool}$ is now taken as the type of the embedded formulas $\varphi_\sigma$, indicating that PML formulas in this minimal embedding now depend only on possible worlds.\footnote{Note that this is also the core idea of the well-known standard translation of PML into first-order logic \cite{BlackburnB07}, which, in fact, is implemented here by exploiting $\lambda$-conversion and compositionality directly in the metalogical HOL without the need for explicit recursive definitions.} Hence, instead of maintaining dependencies on ${W}$, ${R}$, ${V}$, and $w$, the definitions of the logical connectives $a^m$, $\neg^m$, $\supset^m$, and $\Box^m$ now only explicitly maintain a dependency on worlds $w$ (lines 7-10). Their definitions are thus, where $\texttt{V}$ and $\texttt{R}$ are the above-mentioned constant symbols in metalogic HOL: $a^s := \lambda w.\, \texttt{V}\,a\,w$, $\neg^m \varphi := \lambda w.\, \neg (\varphi\,w)$ and $\varphi \supset^m \psi := \lambda w.\, \varphi\,w \longrightarrow \psi\,w$, and $\Box^m \varphi := \lambda w.\, \forall v.\, \texttt{R}\,w\,v \longrightarrow \varphi\,v$. 
The definition of the other logical connectives $\vee^m$, $\wedge^m$, $\Diamond^m$, $\top^m$, and $\bot^m$ (in lines 12-16) is analogous to before. 
Truth in a world $w$, for a given model $\langle {W}, {R}, {V} \rangle$ maintained at the meta-lavel HOL, called \texttt{RelativeTruthM} and noted as $w \models^m \varphi$, is then defined (see line 18) as the application of the embedded formula $\varphi$ to $w$. Consequently, validity $\models^m \varphi$ of an embedded formula $\varphi$ is defined as relative truth w.r.t.~to all worlds $w$ (line 20). 
The bag \texttt{DefM} defined (in lines 22-24) is analogous to \texttt{DefD} and \texttt{DefS} from before. Again, not a single axiom is introduced. So PML is once again characterized as a proper fragment of HOL, but now more lightweight.

\section{Faithfulness automated}
\label{sec:pmlinhol_faithfulness}
\begin{figure}[tp!]
  \centering
  \colorbox{gray!30}{\includegraphics[width=.97\textwidth]{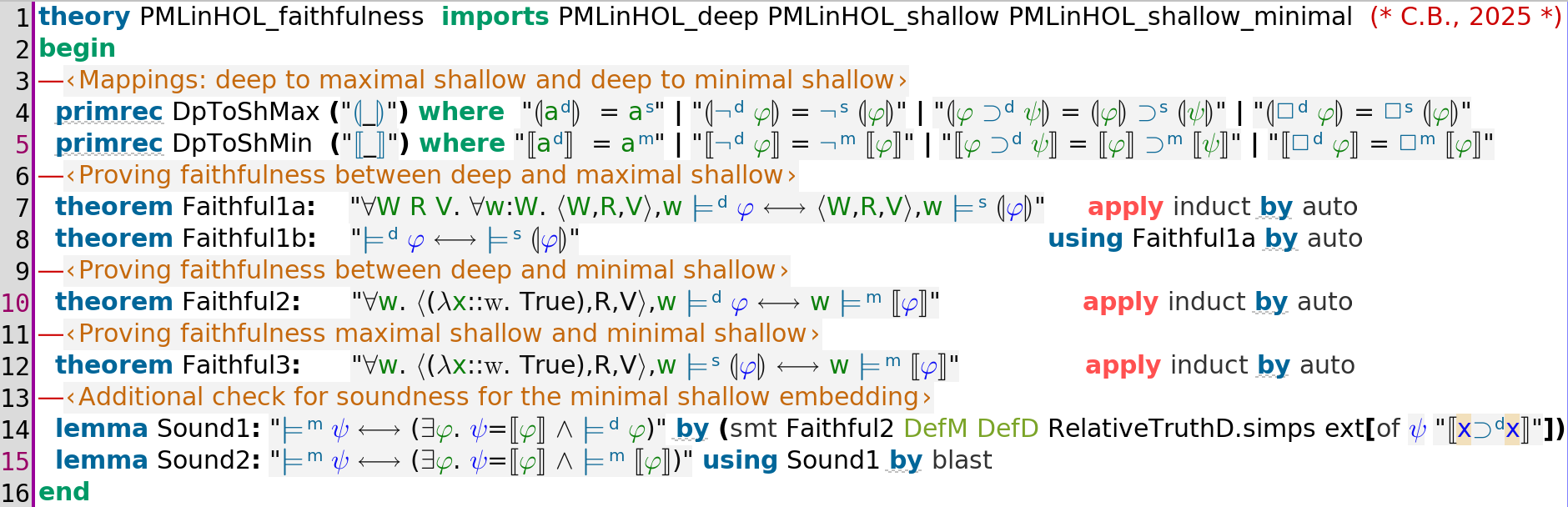}}
  \caption{Automated faithfulness proofs between different embeddings of PML in HOL.}
  \label{fig:PMLinHOL_faithfulness}
\end{figure}

The deep embedding presented in \S\ref{sec:pmlinhol_deep} and the maximal and minimal shallow embeddings presented in \S\ref{sec:pmlinhol_shallow_max} and \S\ref{sec:pmlinhol_shallow_min} have been carefully designed to allow for automated proofs of mutual faithfulness within the metalogic HOL. A relevant aspect is the sharing of the signature $\mathcal{S}$ of propositional symbols between all embeddings. 
The automated faithfulness proofs are shown in Fig.~\ref{fig:PMLinHOL_faithfulness}.

The starting points are the mapping functions $\llparenthesis . \rrparenthesis$ and $\llbracket . \rrbracket$ (in lines 4-5), which map deeply embedded formulas to maximally and minimally embedded counterparts, respectively. Using these mappings the following faithfulness statements, which are formulated relative to the deeply embedded proof language PML, are proved automatically by induction (see lines 7-12):
\begin{align}
    \forall\,W R\,V. \forall w{:}W .\ \langle W,R,V \rangle, w \models^d \varphi & \longleftrightarrow \langle W,R,V \rangle, w \models^s \llparenthesis \varphi \rrparenthesis \tag{\texttt{Faithful1a}} \\
     \models^d \varphi & \longleftrightarrow\ \models^s \llparenthesis \varphi \rrparenthesis \tag{\texttt{Faithful1b}}\\
     \forall\,w.\  \langle (\lambda x_w. \top),\texttt{R},\texttt{V} \rangle, w \models^d \varphi  & \longleftrightarrow  w \models^m \llbracket \varphi \rrbracket \tag{\texttt{Faithful2}} \\
     \forall\,w.\  \langle (\lambda x_w. \top),\texttt{R},\texttt{V} \rangle, w \models^s \llparenthesis \varphi \rrparenthesis & \longleftrightarrow  w \models^m \llbracket \varphi \rrbracket \tag{\texttt{Faithful3}}
\end{align}

Theorems \texttt{Faithful1a} and \texttt{Faithful1b} are rather obvious; they establish the desired faithfulness correspondence between deeply embedded formulas and their maximally shallow counterparts. Theorems \texttt{Faithful2} and \texttt{Faithful3}, on the other hand, require some explanation. The former states that if ${W}$ is taken as $\lambda x_\texttt{w}.\top$, i.e.~the universe of possible worlds denoted by type \texttt{w} in metalogic HOL, and if ${R}$ and ${V}$ are replaced by the metalogical constant symbols $\texttt{R}$ and $\texttt{V}$ in HOL, denoting accessibility between worlds and evaluation of propositional symbols from $\mathcal{S}$ in worlds, then each deeply embedded formula $\varphi$ faithfully corresponds to its minimally shallow counterpart $\llbracket \varphi \rrbracket$. The latter formulates an analogous correspondence between maximal shallow formulas $\llparenthesis \varphi \rrparenthesis$ and minimal shallow formulas $\llbracket \varphi \rrbracket$. Together, these theorems show the intended correspondences between all the introduced embeddings, and they provide a valuable foundation for working with them simultaneously and interchangeably in applications. Moreover, the surjection theorems \texttt{Sound1} and \texttt{Sound2} can be automated by \texttt{sledgehammer} for the minimal shallow embedding, but not yet for the maximal one.  

\section{Experiments}
\label{sec:pmlinhol_experiments}

\begin{figure}[btp!]
  \centering
  \colorbox{gray!30}{\includegraphics[width=.97\textwidth]{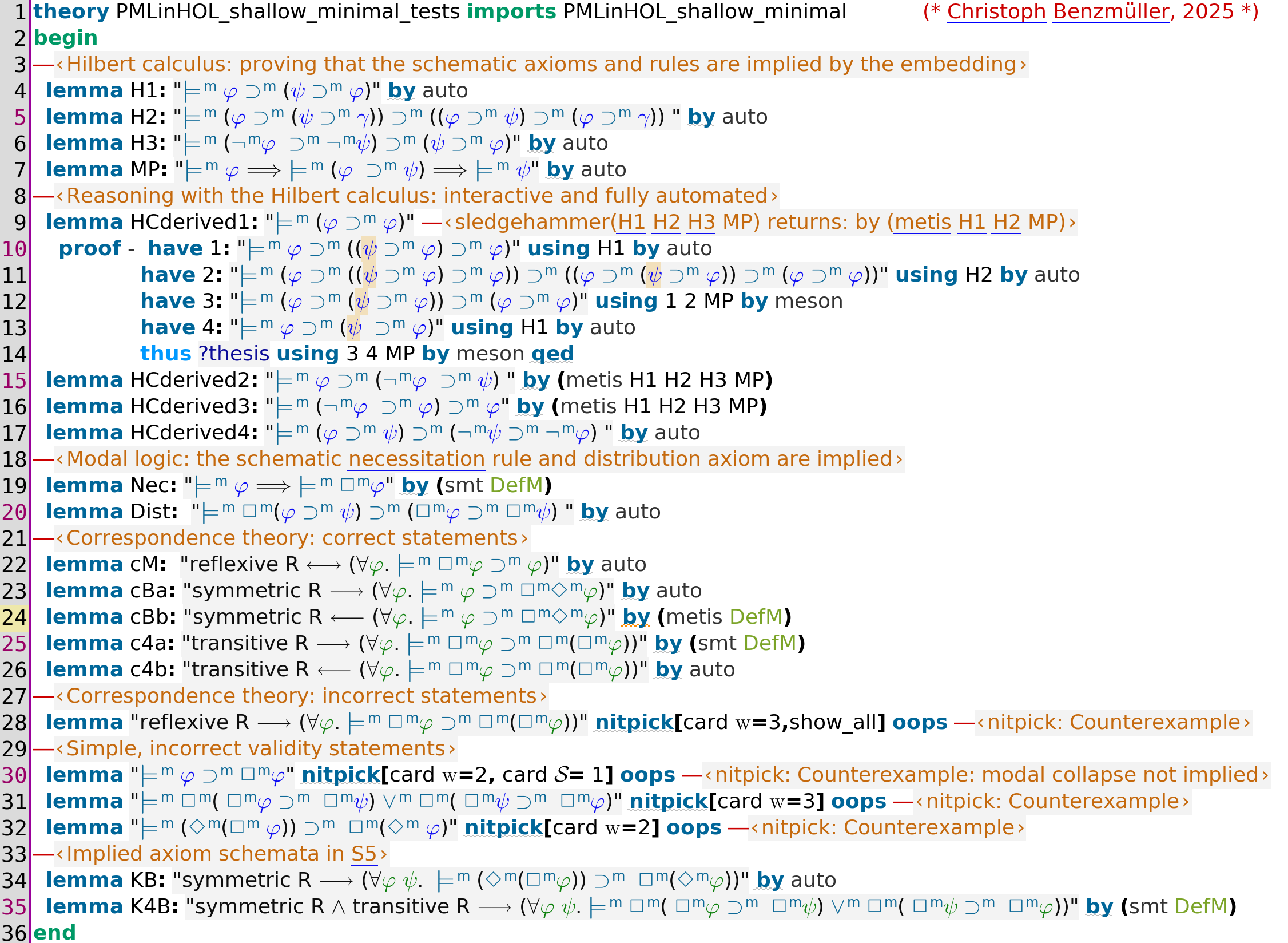}}
  \caption{Experiments: Hilbert system and Sahlqvist correspondences.}
  \label{fig:PMLinHOL_shallow_minimal_tests}
\end{figure}

The possibilities that the linked embeddings offer, e.g., for logic education, are illustrated with some experiments.
The experiments performed also test the soundness of our embeddings, and assess proof automation and counterexample finding for them with \texttt{sledgehammer} and \texttt{nitpick}. The embedding that has performed best in this regard is the minimal shallow embedding. This is why file \texttt{PMLinHOL\_shallow\_minimal\_tests.thy}, see Fig.~\ref{fig:PMLinHOL_shallow_minimal_tests}, is presented here.\footnote{Corresponding files for the deep and the maximal shallow embeddings are shown in the appendix. %of \cite{R94}. 
The Isabelle/HOL sources are available at \url{http://logikey.org/tree/master/CoursesAndTutorials/2025-CADE-DeepShallow}.}

\paragraph{Derived Hilbert calculus.} The first experiments (see lines 4-7) confirm that the schematic axioms \texttt{H1}-\texttt{H3} and the modus ponens rule \texttt{MP} of the propositional Hilbert calculus from \S\ref{sec:pml} are implied by the minimal shallow embedding (local versions of \texttt{H1}-\texttt{H3} and \texttt{MP} rule are also implied; this is not shown here). 
The \texttt{auto} tactic, and with it, the simplifier \texttt{simp} are used, which has access to the definitions in bundle \texttt{DefM}.
The $\varphi$, $\psi$ and $\gamma$ are schematic variables of type $\sigma$ in the metalogic HOL. The proofs were originally found by calling \texttt{sledgehammer}. 

\paragraph{Hilbert style proofs automated and interactive.}
The next experiment (in lines 9-14) demonstrates automated and interactive reasoning using the schematic axioms and rules of the Hilbert calculus introduced earlier. It is shown that the formula schema $\varphi \supset^m \varphi$ is derivable in the given Hilbert calculus. By preventing access to the definitions in the bag \texttt{DefM}, e.g.~with the command \texttt{\texttt{sledgehammer}(H1 H2 H3 MP)}, the automated theorem provers orchestrated by \texttt{sledgehammer} can be forced to search for proofs in the given Hilbert calculus only, and they succeed here. They find out that only \texttt{H1} and \texttt{H2} and \texttt{MP} are needed to obtain the result. In the given situation, the reconstruction of a proof from \texttt{H1} and \texttt{H2} and \texttt{MP} is then possible with Isabelle's trusted \texttt{metis} tool.
For educational purposes it is relevant that proofs from the literature can be easily mapped one to one to proofs in our setting. This is illustrated here with an interactive derivation (in lines 10-14) that is following a proof from a Wikipedia entry \cite{HilbertSystem} on Hilbert systems.
Further derived axiom schemata are proved automatically  at the level of the Hilbert calculus (in lines 15-16). If such attempts fail (as was initially the case in line 17), then either interactive proof is needed or, as is demonstrated here with the call of \texttt{auto}, the unfolding of the definitions of the shallow embedding in metalogic HOL (within \texttt{auto}) may in fact help to still find a proof automatically. Depending on prior knowledge and teaching objectives, students can be asked to solely provide interactive proofs in such exercises rather than using automation tools. It is also important to note that Isabelle/HOL's proof tactics support, and in particular the Eisbach package, can be used to further enhance the automation of proofs at the level of embedded logic.\footnote{The Eisbach package was e.g.~used in \cite{KirchnerPhD,J50} to define and automate proof methods for a non-trivial shallow embedding of abstract object theory AOT \cite{PLM} in HOL.}

\paragraph{Correspondence theory automated.} Modal logic textbooks typically include exercises on the well known Sahlqvist correspondences \cite{Sahlqvist1975} (see lines 22-26). Using the definitions in bag \texttt{DefM}, such correspondences are here automated in Isabelle/HOL. More exhaustive experiments on such (and similar) correspondences have been reported in prior work~\cite{B12,C47,IntuitionisticCube,J68}.

\begin{figure}[tp!]
\begin{minipage}{.5\textwidth} \centering
\begin{tikzpicture}[state/.style=state with output,shorten >=1pt,node distance=2cm,on grid,auto] 
   \node[state] (i_1)   {$w_1$ \nodepart{lower} $\varphi$}; 
   \node[state] (i_2) [right=of i_1] {$w_2$ \nodepart{lower} $\neg \varphi$}; 
   \node[state,initial below] (i_3) [right=of i_2] {$w_3$\nodepart{lower} $\varphi$}; 
    \path[->] 
    (i_1) edge [loop above] node {} ()
          edge [bend right] node {} (i_2) 
          edge [bend right] node {} (i_3) 
    (i_2) edge [loop right] node {} () 
    (i_3) edge [loop above] node {} () 
          edge [bend right] node {} (i_1);
\end{tikzpicture}
\end{minipage}
\begin{minipage}{.47\textwidth} 
\includegraphics[width=.95\textwidth]{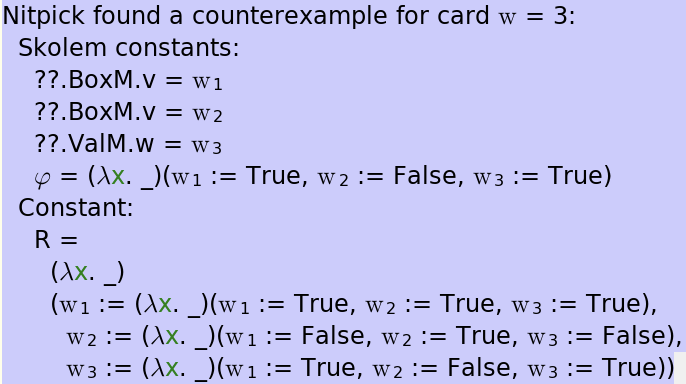}
\end{minipage}
\caption{Counterexample reported by \texttt{nitpick} in line 28 of Fig.~\ref{fig:PMLinHOL_shallow_minimal_tests}}
 \label{fig:Counterexample}
\end{figure}

\paragraph{Disproving correspondences.} Even more interesting than automatically proving the Sahlqvist  correspondences is disproving falsely assumed correspondences. For example, when \texttt{nitpick} is called on the false conjecture, that in PML models with reflexive accessibility relations $R$ the axiom schema $\Box^m \varphi \supset^m \Box^m  \Box^m \varphi$ is implied (see line 28), the counterexample shown in Fig.~\ref{fig:Counterexample} is reported back. 

As prior work demonstrates, see e.g.~\cite{C47,J40,J68}, \texttt{nitpick} often provides intuitive and sometimes even surprising counterexamples when used in conjunction with shallow embeddings.\footnote{Prior work~\cite{C47,IntuitionisticCube} also illustrates how counterexamples reported by \texttt{nitpick} can be systematically converted into statements that verifiably disprove the false conjectures; the method presented there could indeed be fully automated.}  This is not only useful for teaching logic, but can also be useful for research studies, see e.g.~\cite{J40}. However, some non-classical logics have only infinite models, and then the finite model finder \texttt{nitpick} is no longer applicable; this is e.g. the case for Lorini's T-STIT logic \cite{Lorini2013}, which was embedded in HOL by Meder \cite{MederMasters}.
Developing infinite model finding tools and integrating them into proof assistants such as Isabelle/HOL is thus an important research challenge. So too is the automated, domain-specific provision of intuitive counterexample diagrams like the one on the left in Fig.~\ref{fig:Counterexample}.

For further false conjectures (in lines 30-32 of Fig.~\ref{fig:PMLinHOL_shallow_minimal_tests}), including the modal collapse formula $\varphi \supset^m \Box^m \varphi$,  intuitive counterexamples are generated by \texttt{nitpick}. Finally, examples of simple modal theorems (in lines 34-35), such as the KB theorem $\Diamond^m \Box^m  \varphi \supset^m \Box^m \Diamond^m \varphi$, are automated efficiently using \texttt{smt} solvers in combination with definition unfolding in metalogic HOL.

\begin{figure}[tp!]
  \centering
  \colorbox{gray!30}{\includegraphics[width=.97\textwidth]{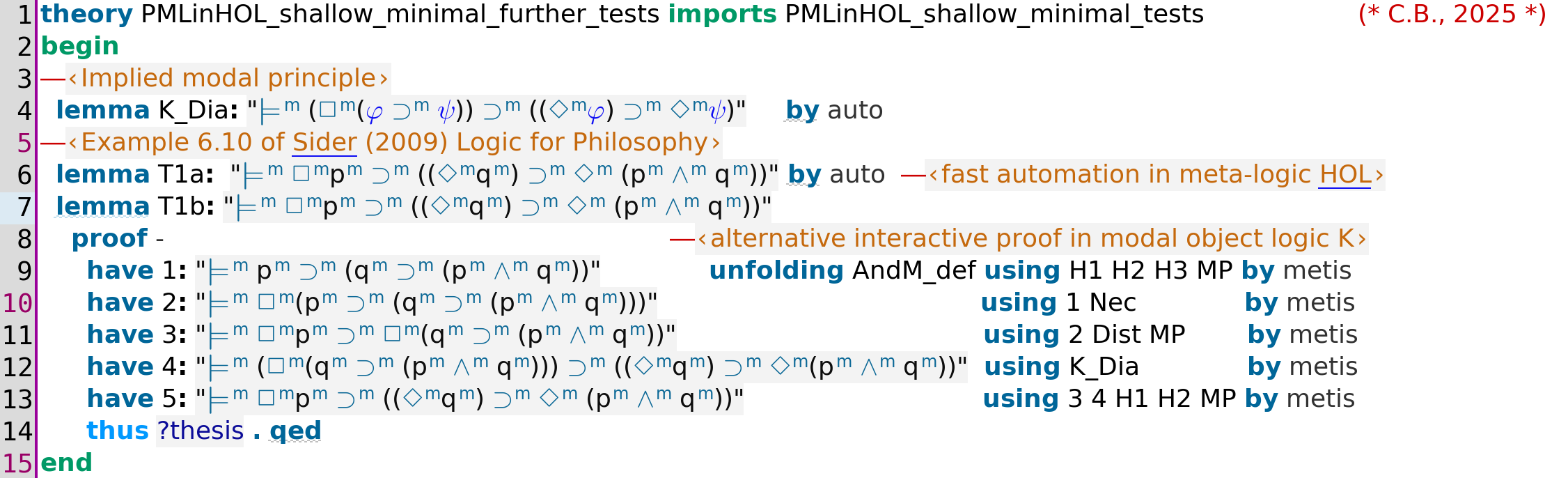}}
  \caption{Experiments: Simple proof in modal logic \textbf{K}.}
  \label{fig:PMLinHOL_shallow_minimal_further_tests}
\end{figure}

\paragraph{Reasoning at object level.} 
Figure \ref{fig:PMLinHOL_shallow_minimal_further_tests} presents 
further experiments. After proving the schematic modal principle $K\Diamond$ (in line 4) an object level proof example \cite[Ex.~6.10]{Sider2009} is studied (in lines 6-14): $\Box^m p^m \supset^m (\Diamond^m q^m \supset^m \Diamond^m (p^m \wedge^m q^m))$, where $p$ and $q$ are propositional symbols from the PML signature $\mathcal{S}$. This statement is again effectively automated via definition unfolding at the metalevel HOL (line 6), while the presented interactive proof (in lines 8-14) uses the propositional Hilbert calculus from before in combination with the modal rule of necessitation, the distribution axiom of base logic $\mathbf{K}$ and the newly introduced \textbf{$K\Diamond$}-principle.

Larger object-level proof examples could now be presented, studied and automated using the minimal shallow embedding. Prior studies have shown that this approach is generally competitive, in particular, when it comes to automating quantified modal logics; see e.g.~the recent experiments reported in \cite{J72}. Recipes for embedding modal quantifiers and extensive application studies for quantified modal logics have been presented in prior works; see e.g.~\cite{J23,J75}.

\begin{figure}[tp!]
  \centering
  \colorbox{gray!30}{\includegraphics[width=.97\textwidth]{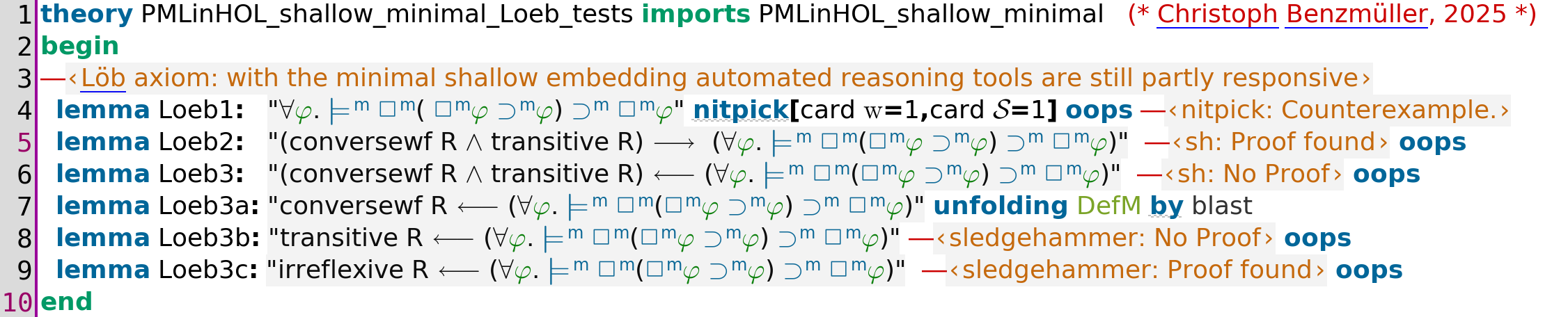}}
  \caption{Experiments: Studying the Löb-axiom of provability logic.}
  \label{fig:PMLinHOL_shallow_minimal_Loeb_tests}
\end{figure}

\paragraph{Experiments with the Löb-axiom.} 
Figure~\ref{fig:PMLinHOL_shallow_minimal_Loeb_tests} presents automation experiments with the Löb-axiom of provability logic \cite{Boolos1994}. 
A first observation, interesting for education and research, 
is that via definition unfolding the schematic axioms are  automatically converted into corresponding formulas in metalogic HOL. For the Löb-axiom (in line 4) we thus obtain after definition unfolding the HOL formula 
$\forall \varphi\, w.\, (\forall v.\, \texttt{R} w v \longrightarrow (\forall z.\, \texttt{R} v z \longrightarrow \varphi z) \longrightarrow \varphi v) \longrightarrow (\forall v.\, \texttt{R} w v \longrightarrow \varphi v)$.
When using the maximal shallow embedding instead, the Löb-axiom unfolds into the expected much heavier HOL formula $\forall\varphi\, W R\, V\, x.\, W\, x \longrightarrow (\forall z.\, W\, z \longrightarrow R x z \longrightarrow (\forall x.\, W x \longrightarrow R z x \longrightarrow \varphi\, W R V x) \longrightarrow \varphi\, W R V z) \longrightarrow (\forall z.\, W z \longrightarrow R x z \longrightarrow \varphi\, W R V z)$, which also explains why automated reasoning is generally much harder for maximal shallow embeddings than for minimal shallow embeddings. The utilisation of the features of the presented framework enables students and researchers to explore frame properties for non-trivial properties independently and to subsequently simplify them. Proof automation or interactive proof can then be employed to investigate their equivalence to already known properties, as illustrated in Fig.~\ref{fig:PMLinHOL_shallow_minimal_Loeb_tests} (in lines 5-9). As shown, such proof automation attempts succeed for the following conjectures: (line 5) the converse well-foundedness of the accessibility relation $\texttt{R}$ in combination with the transitivity of $\texttt{R}$ implies the Löb axiom, (line 7) the Löb axiom implies converse well-foundedness of  $\texttt{R}$, and (line 9) the Löb axiom implies the irreflexivity of $\texttt{R}$. Whenever transitivity is an implicant to be proved, proof automation still fails (see lines 6 and 8). Here interactive proof is still required or, alternatively, improved HOL theorem proving systems, in particular provers that can suitably guess non-trivial instantiations for higher-order order variables.\footnote{For more on this challenge topic see e.g.~the experiments reported in~\cite{J63}.}

\paragraph{Performance: examples so far.} The performance of $\texttt{sledgehammer}$ and $\texttt{nitpick}$ for the different embeddings for the example proof problems as discussed so far is briefly discussed:\footnote{\label{foot:MacBook}A MacBook Pro 16,1 (6-Core Intel Core i7 processor, 2,6 GHz, 6 Cores, 256 KB L2 Cache per Core, 12 MB L3 Cache, 16 GB Memory) and Isabelle 2025 in standard configuration was used, which sets a 30s prover timeout for $\texttt{sledgehammer}$.}
\emph{Minimal shallow embedding:} Proofs for 36 out of 38 valid statements (theorems, lemmas and substatements in interactive proofs) 
were found by calls to $\texttt{sledgehammer}$.
\texttt{Nitpick} reported counterexamples for 5 out of 5 invalid statements. Reconstruction of $\texttt{sledgehammer}$ proofs with trusted tactics failed in 2 cases. 
\emph{Maximal shallow embedding:} Proofs for 34 out of 38 valid statements were found by \texttt{sledgehammer}.
Countermodels for 4 out of 5 invalid statements were found by \texttt{nitpick}. Reconstruction of \texttt{sledgehammer} proofs with trusted tactics failed in 4 cases. 
\emph{Deep embedding:} Proofs for 32 out of 38 valid statements were found by $\texttt{sledgehammer}$; in particular, there were no proofs delivered for the Löb examples.
Countermodels for 5 out of 5 invalid statements were found by nitpick. Reconstruction of proofs failed in 2 cases. 

\paragraph{Performance: formula classification.} 
Further experiments were carried out in order to assess the performance of the different embeddings more rigorously. 
These tests focused on the classification of PML formula examples with respect to the logics in the modal logic cube \cite{sep-logic-modal} defined by the schematic axioms $T:= \Box \varphi \supset \varphi$,  $B:= \varphi \supset \Box \Diamond\varphi$ and $4:= \Box \varphi \supset \Box \Box \varphi$, respectively by their semantic counterparts $Ts := \texttt{reflexiv}\,R$, $Bs := \texttt{symmetric}\,R$ and $4s := \texttt{transitiv}\,R$ (with $R$ being the accessibility relation). The example formulas studied were:
$$\begin{array}{rl@{\qquad}rl}
F_1{:} & \Diamond\Diamond \varphi \supset  \Diamond \varphi  &
F_6{:} & ((\Box(\varphi \supset \psi)) \wedge \Diamond\Box\neg\psi) \supset \neg\Diamond\psi \\
F_2{:} & \Diamond\Box \varphi \supset \Box\Diamond \varphi &
F_7{:} & \Diamond \varphi \supset \Box(\varphi \vee  \Diamond \varphi) \\
F_3{:} & \Diamond\Box \varphi \supset \Box \varphi &
F_8{:} & \Diamond(\Box \varphi) \supset (\varphi \vee  \Diamond \varphi)\\
F_4{:} & \Box\Diamond\Box\Diamond \varphi \supset \Box\Diamond \varphi  &
F_9{:} & (\Box\Diamond \varphi \wedge \Box\Diamond\neg \varphi) \supset  \Diamond\Diamond \varphi \\
F_5{:} & \Diamond(\varphi \wedge \Diamond\psi)) \supset (\Diamond \varphi \wedge \Diamond\psi) &
F_{10}{:} & (\Box(\varphi \supset \Box\psi) \wedge \Box\Diamond\neg\psi) \supset \neg\Box\psi
\end{array}$$
The automatically identified weakest logics in the modal cube in which these formulas are valid are as follows: 
$F_1{:}$ $\textbf{K4}$,
$F_2{:}$ $\textbf{KB}$,
$F_3{:}$ $\textbf{KB4}$,
$F_4{:}$ $\textbf{KB/K4}$,
$F_5{:}$ $\textbf{K4}$,
$F_6{:}$ $\textbf{KB4}$,
$F_7{:}$ $\textbf{KB4}$, 
$F_8{:}$ $\textbf{KT/KB}$,
$F_9{:}$ $\textbf{KT}$, and
$F_{10}{:}$ $\textbf{KT}$.

Explained in abstract terms, for all $M \subseteq \{T,B,4\}$, $N \subseteq \{Ts,Bs,4s\}$ and $F \in \{F_1,\ldots,F_{10}\}$, \texttt{sledgehammer} and \texttt{nitpick} were called to assess the validity of the following sets of proof problems (each one consisting of $3^2\times 10 = 80$ entries): $M \models^d F$, $N \models^d F$, $M \models^s F$, $N \models^s F$, $M \models^m F$, $N \models^m F$. 
\texttt{sledgehammer} and \texttt{nitpick} were thereby invoked in two modes, 
with and without prior application of Isabelle/HOL's simplifier \texttt{simp} to the proof problem under consideration. Consequently, for each set of problems, a total of 160 calls to each tool were conducted (80 with prior simplification and 80 without).

The results of these experiments, conducted with the same hardware as mentioned in Footnote \ref{foot:MacBook}, are presented in Fig.~\ref{fig:experimentresults}.
\begin{figure}[t]
    \centering
\begin{tabular}{c|ccc|cc|c}
\multicolumn{1}{c}{}  & \multicolumn{3}{c}{\texttt{nitpick}} & \multicolumn{2}{c}{\texttt{sledgehammer}} & \multicolumn{1}{c}{} \\
      & $A$ & $B$ & $C$ & $D$ & $E$ & %$A{+}B{+}D$ 
      score \\ 
Problems         & \,counterex.\, & \,no counterex.\, & \,unknown\, & \,proof found\, & \,no proof\, & \\ \hline
$M \models^m F$ &  \textbf{84(42/42)} & 72(36/36) & 4(2/2) & 73(\textbf{38}/35) & 87(\textbf{42}/45) & 229 \\
$N \models^m F$ & \textbf{84(42/42)} & \textbf{76(38/38)} & \textbf{0(0/0)} & \textbf{76(38/38)} & \textbf{84(42/42)} & \textbf{236} \\
$M \models^s F$ &  32(16/16) & 0(0/0) & 128(64/64) & 32(24/8) & 128(56/72) & 64  \\
$N \models^s F$ & \textbf{84(42/42)} &  0(0/0) & 76(38/38) & 66(\textbf{38}/28) & 94(\textbf{42}/52) & 150 \\
$M \models^d F$ &  16(10/6) & 74(0/74) & 70(70/\textbf{0}) & 33(16/17) & 127(64/63) & 123 \\
$N \models^d F$ & 32(8/24) & 56(0/56) & 72(72/\textbf{0}) & 70(\textbf{38}/32) & 90(\textbf{42}/48) & 158 \\
\end{tabular}
    \caption{Performance results of the automated PML formula classification task conducted with \texttt{nitpick} and \texttt{sledgehammer} (explanation: the first entry $73(\textbf{38}/35)$ in column $D$, e.g.,~states that 73 proofs were reported in total by \texttt{sledgehammer} for the problem set $M \models^m F$, with 38 proofs found with prior application of the simplifier \texttt{simp} and 35 without; analogous for the other entries).}
    \label{fig:experimentresults}
\end{figure}
The best results are highlighted. These are the maxima for the informative cases `counterexample found' and `no (finite) counterexample found' for \texttt{nitpick},
and `proof found' for \texttt{sledgehammer}. For the 'unknown' and 'no proof' cases, which don't provide any useful information, the minima are shown as the best results. A overall score is reported, which was computed by adding the results obtained for the informative cases.

The results obtained are briefly discussed. Perfect experimentation results were obtained for the problem set $N \models^m F$, that is, the minimal shallow embedding when used in combination with the semantical frame conditions $Ts$, $Bs$ and $4s$. 
In this setting \texttt{nitpick} either provided a counterexample for each formula classification task or reported that there was no (finite) counterexample. The latter judgments were then supplemented by proofs from \texttt{sledgehammer} that confirmed the validity of these formulas. The second best performance was achieved with the minimal shallow embedding when used in combination with the axiom schemata $T$, $B$ and $4$. 
The maximal shallow embedding returned some good results for the problem set $N \models^s F$ regarding counterexample finding with \texttt{nitpick} and regarding theorem proving with \texttt{sledgehammer} when prior simplification was applied. However, in this setting, \texttt{nitpick} was unable to identify any formulas with no (finite) countermodels.
For the deep embedding the performance results obtained with \texttt{nitpick} regarding counterexample finding significantly dropped in comparison to the minimal shallow embedding. The numbers reported for the deep embedding regarding the identification of formulas with no (finite) countermodels are still good though.\footnote{However, some seemingly self-contradictory answers by \texttt{nitpick} in this setting with and without prior simplification for some problems require further consideration.} Interestingly, the maximal number of 38 proofs was still obtained by \texttt{sledgehammer} in the deep embedding when prior simplification was applied. Future work should extend the experiments presented to more challenging test formulas $F$ and also include the further axiom schemata (and their corresponding semantic conditions) from the modal logic cube \cite{sep-logic-modal}. Additionally, it seems useful to replay, now using all the embeddings as provided in this paper, the metalogical cube verification experiments reported in prior work \cite{B12,C47}.

\section{Related work and implications for \logikey}
\label{sec:logikey}

The notions of deep and shallow embeddings appear to originate from the work of Boulton et al.~\cite{BoultonGGHHT92}, and the connections between the two have been discussed in numerous related papers, see e.g.~\cite{SvenningssonA12,KaposiKK19,PrinzKL22,WangCT24} and the references therein.  However, to the best of the author's knowledge, no previous work has succeeded in integrating deep and (various degrees of) shallow embeddings in the form presented and, in particular, in enabling mutual faithfulness proofs between them.

\paragraph{Deep embeddings.}
Exemplary developments and applications of non-trivial deep logic embeddings include self-verifications of the HOL light logic \cite{Harrison06,AbrahamssonMKS22}, a formalization of the Core Why3 logic \cite{10.1145/3632902}, and a provably sound implementation of a differential dynamic logic \cite{deep2017}, to name a few. For examples of deep embeddings of logics and their calculi in the Coq proof assistant \cite{Coq}, see e.g.~\cite{SozeauABCFKMTW20} and the references therein, and Isabelle/HOL examples are presented in \cite{BlanchettePT17,FOL_Seq_Calc3-AFP,Implicational_Logic-AFP,LIPIcs.ITP.2022.13}. Further related work on the self-verification or bootstrapping of theorem provers 
can be found in \cite{Carneiro20}, and more recently in \cite{carneiro2024lean4leanverifiedtypecheckerlean} in the context of Lean \cite{Lean}.

\paragraph{Shallow embeddings.}
Exemplary work in this category, which Nipkow called the extensional approach \cite{Nipkow2002}, includes a mechanization of HOL reasoning in first-order set theory \cite{GuilloudGGK24}, the techniques presented in~\cite{KaposiKK19} to overcome some problems of shallow embedding for reasoning over the syntax of embedded languages, a shallow embedding for pure type systems in first-order logic~\cite{LIPIcs.TYPES.2016.9}, and various other works such as~\cite{RabeLP14,LammichM12}; see also the references given in \cite{WangCT24}.
Note also that the automation support for various quantified non-classical logics as realized in the HOL theorem prover Leo3 \cite{J51,J72} is based on minimal shallow embeddings.

\paragraph{Implications for {LogiKEy}.}
The logic-pluralistic knowledge and reasoning methodology \logikey \cite{J48}, see Fig.~\ref{fig:LogiKEy}, leverages higher-order logic (HOL) as a unifying metalogic to enable the modeling of diverse object logics, so far,  primarily via using shallow semantic embeddings. \logikey\ has been used in a range of application domains and contexts. Very different kinds of object logics have been embedded \logikey\ over the past decade, including, for example, free logic \cite{J40}, modal higher-order logic \cite{J75}, public announcement logic \cite{J58}, conditional logic \cite{J31} and dyadic deontic logic with a preference-based semantics \cite{J68}.

This paper shows how the \logikey\ approach could benefit in the future from providing faithful, deep and shallow embeddings of all object logics under consideration. This would allow the embeddings to be used interchangeably in applications.
The technique presented has also the potential to replace or complement pen-and-paper faithfulness proofs in prior works, such as the ones reported in \cite{J23,J31,J58}, by computer-verified proofs in Isabelle/HOL. This in turn will be relevant for enabling applications in areas that require highest levels of trust. For this note that a particular application direction of \logikey\ is normative and legal reasoning in the context of AI regulation, see also \cite{C101,J48}.

This paper also illustrates the conceptual modifications applied in prior \logikey\ papers when transitioning from lightweight shallow embeddings of non-classical logics, such as those used to encode higher-order modal logic \cite{J23} or conditional logic \cite{J31}, which consider only dependencies on possible worlds, to more heavyweight shallow embeddings, as, for example, required to embed public announcement logic \cite{J58}; in the latter work additional dependencies on evaluation domains had to be encoded. Note on the other hand, that previous work has often avoided the explicit modeling of atoms, signatures, and evaluations, which has been essential for the work presented in this paper. Moreover, the presented technique should eventually help to overcome challenges in application contexts that require the combined use of deep and shallow embeddings. The Fatio protocol for multi-agent argumentation \cite{W72} is such an example.
Most importantly, the presented technique for linking deep and shallow embeddings will be very fruitful for furthering the use of \logikey\ in university-level logic courses, including the author's courses on universal logical reasoning that have been taught for many years at FU Berlin and now at the University of Bamberg.

\tikzset{block/.style={draw, thick, text width=2cm, minimum height=1.3cm, align=center}, 
         line/.style={-latex} }
\tikzset{ font={\fontsize{12pt}{12}\selectfont}}

\tikzset{testpic1/.pic={ 
\node[block, fill=green!20, text width=10cm, minimum width=14cm, minimum height=2cm] (m1) 
      { \textbf{\huge L3 --- Applications} };
\node[block, below=.5cm of m1, fill=orange!30, text width=14cm, minimum height=2cm] (m2) 
      { \textbf{\huge L2 --- Domain-Specific Language(s)} };
\node[block, below=.5cm of m2, fill=yellow!20, text width=10cm, minimum width=14cm, minimum height=2cm] (m3) 
      { \textbf{\huge L1 --- Object Logic(s)} };
\node[block, below=.5cm of m3, fill=blue!20, text width=11cm, minimum width=14cm, minimum height=2cm] (m4) 
      { \textbf{\huge L0 --- Meta-Logic (HOL)} };
\node[block, above right=-.5cm and 1.5cm of m1, fill=gray!40, text width=8cm, minimum width=9cm, minimum height=2cm, rotate=-90] (r1) 
      { \textbf{\huge \logikey\ Methodology} };

  \begin{scope}[on background layer] 
    \node[draw, fill=blue!20, inner xsep=5mm, inner ysep=5mm, fill opacity=0.6, fit=(m1)(m2)(m3)(m4)](m1tom4){};
  \end{scope}

  \begin{scope}[on background layer] 
    \node[draw, fill=gray!40, inner xsep=20mm, inner ysep=5mm, fill opacity=0.5, fit=(m1)(m2)(m3)(m4)(m1tom4)](all){};
  \end{scope}
}}

\tikzset{testpic2/.pic={ 
\node[block, fill=green!20, text width=10cm, minimum width=14cm, minimum height=2cm] (m1) 
      { \mbox{\textbf{\huge Gödel's Ontological Argument}} };
\node[block, below=.5cm of m1, fill=orange!30, text width=14cm, minimum height=2cm] (m2) 
      { \textbf{\huge Theory of Positive Properties} };
\node[block, below=.5cm of m2, fill=yellow!20, text width=10cm, minimum width=14cm, minimum height=2cm] (m3) 
      { \textbf{\huge Higher-Order Modal Logic} };
\node[block, below=.5cm of m3, fill=blue!20, text width=11cm, minimum width=14cm, minimum height=2cm] (m4) 
      { \textbf{\huge Meta-Logic (HOL)} };

  \begin{scope}[on background layer] 
    \node[draw, fill=blue!20, inner xsep=5mm, inner ysep=5mm, fill opacity=0.6, fit=(m1)(m2)(m3)(m4)](m1tom4){};
  \end{scope}
}}

\tikzset{testpic3/.pic={ 
\node[block, fill=green!20, text width=10cm, minimum width=14cm, minimum height=2cm] (m1) 
      { \mbox{\textbf{\huge Wise Men Puzzle}} };
\node[block, below=.5cm of m1, fill=orange!30, text width=14cm, minimum height=2cm] (m2) 
      { \textbf{\huge Theory of Agent(s) Knowledge} };
\node[block, below=.5cm of m2, fill=yellow!20, text width=10cm, minimum width=14cm, minimum height=2cm] (m3) 
      { \textbf{\huge Public Announcement Logic} };
\node[block, below=.5cm of m3, fill=blue!20, text width=11cm, minimum width=14cm, minimum height=2cm] (m4) 
      { \textbf{\huge Meta-Logic (HOL)} };

  \begin{scope}[on background layer] 
    \node[draw, fill=blue!20, inner xsep=5mm, inner ysep=5mm, fill opacity=0.6, fit=(m1)(m2)(m3)(m4)](m1tom4){};
  \end{scope}
}}

\begin{figure}[tp!]
\centering 
\begin{minipage}{.70\textwidth}
\resizebox{\textwidth}{!}{
    \begin{tikzpicture}
      \pic{testpic1};
    \end{tikzpicture}
    }
\end{minipage}
\hfill
\begin{minipage}{.28\textwidth}
\resizebox{\textwidth}{!}{
    \begin{tikzpicture}
      \pic{testpic2};
    \end{tikzpicture}
    }
\vskip.3em
\resizebox{\textwidth}{!}{    
    \begin{tikzpicture}
      \pic{testpic3};
    \end{tikzpicture} 
    }
\end{minipage}
\caption{The logic-pluralistic knowledge representation and reasoning methodology \logikey (left) and two concrete applications of it (right): Studies on Gödel's ontological argument \cite{J75} (top) and an automation of the wise men puzzle \cite{J58} (bottom). \label{fig:LogiKEy}}
\end{figure}

\section{Conclusion}
\label{sec:conclusion}
This paper has demonstrated how the simultaneous deployment of deep logic embeddings and shallow logic embeddings of various
degrees in classical higher-order logic can be achieved. This technique allows for the mechanization of faithfulness proofs between the interlinked embeddings, and, as has been shown in this paper, this technique can even support the automation of such proofs. 

The work presented is relevant to future research in various areas. Due to its conciseness and elegance, as well as its support for interactive and automated theorem proving and counterexample finding, it can also be used to support logic education effectively. 
The latter application direction was the main motivation for illustrating this approach using simple propositional modal logic in this paper. This logic is routinely used as introductory material in logic courses worldwide. The Isabelle/HOL sources provided and explained in this paper should be easily reusable to support such courses, and they should also be easily transferable to other higher-order proof assistant systems.

\subsubsection{Acknowledgements}
I am grateful to the reviewers of this paper for their valuable feedback. I would also like to thank D.~Fuenmayor, D.~Kirchner, L.~Pasetto and E.~Zalta for their comments and all those who have contributed to \logikey\ in the past. My thanks also go to Deepl for the fruitful exchange that (hopefully) helped to improve the quality of the wording of this paper.

\bibliographystyle{splncs04}
\bibliography{chris,literature}
%\end{document}

\begin{appendix}

\begin{comment}
\section{Preliminaries}
\label{sec:preliminaries}
Figure~\ref{fig:Preliminaries} presents preliminaries that are uniform for all embeddings studied and which are imported by all Isabelle/HOL theory files presented in this paper. 

Two HOL base types $\texttt{w}$ and $\mathcal{S}$ are declared (in lines 4-5) to denote a (nonempty) set of possible worlds and a (nonempty) set of propositional constant symbols, the signature common to all embeddings discussed in this paper. Then (in line 6) the exemplary signature  symbol $\texttt{p}$, 
%$\texttt{p}$ and $\texttt{r}$ 
is introduced, and type synonyms are declared (in lines 7-8): $\mathcal{W}$ for sets of possible worlds (universe), $\mathcal{R}$ for binary relations on $W$ (accessibility relations), and
$\mathcal{V}$ for evaluation functions that assign truth values to the symbols $a \in \mathcal{S}$ for each world $w$. Then some well-known relational properties are introduced as abbreviations (in lines 11-19), and the guarded quantifier notation $\forall w{:}W.\, \varphi$ is defined to stand for $\forall w. \,W\,w \longrightarrow \varphi$ ($W$ is of type $\mathcal{W}$). Finally, the useful backward implication $\longleftarrow$ is defined, which is unfortunately missing in Isabelle/HOL.

\begin{figure}[!ht]
  \centering
  \colorbox{gray!30}{\includegraphics[width=.97\textwidth]{isabelle/PMLinHOL_preliminaries.png}}
  \caption{Preliminaries.}
  \label{fig:Preliminaries}
\end{figure}
\end{comment}

\section{Alternative experiments}
\label{sec:alternative_experiments}
\begin{figure}[htp!]
  \centering
  \colorbox{gray!30}{\includegraphics[width=.97\textwidth]{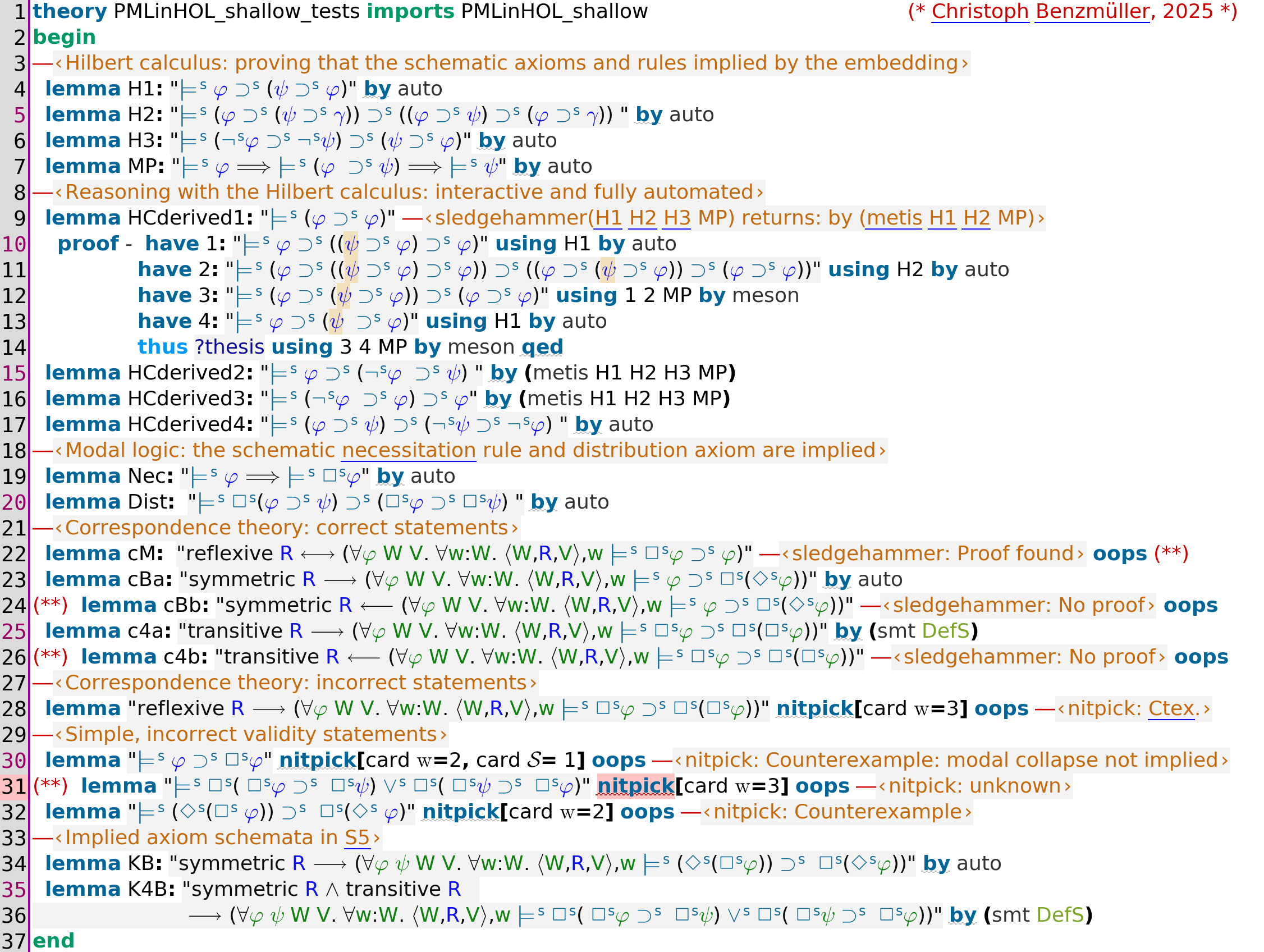}}
   \vskip.5em
  \colorbox{gray!30}{\includegraphics[width=.97\textwidth]{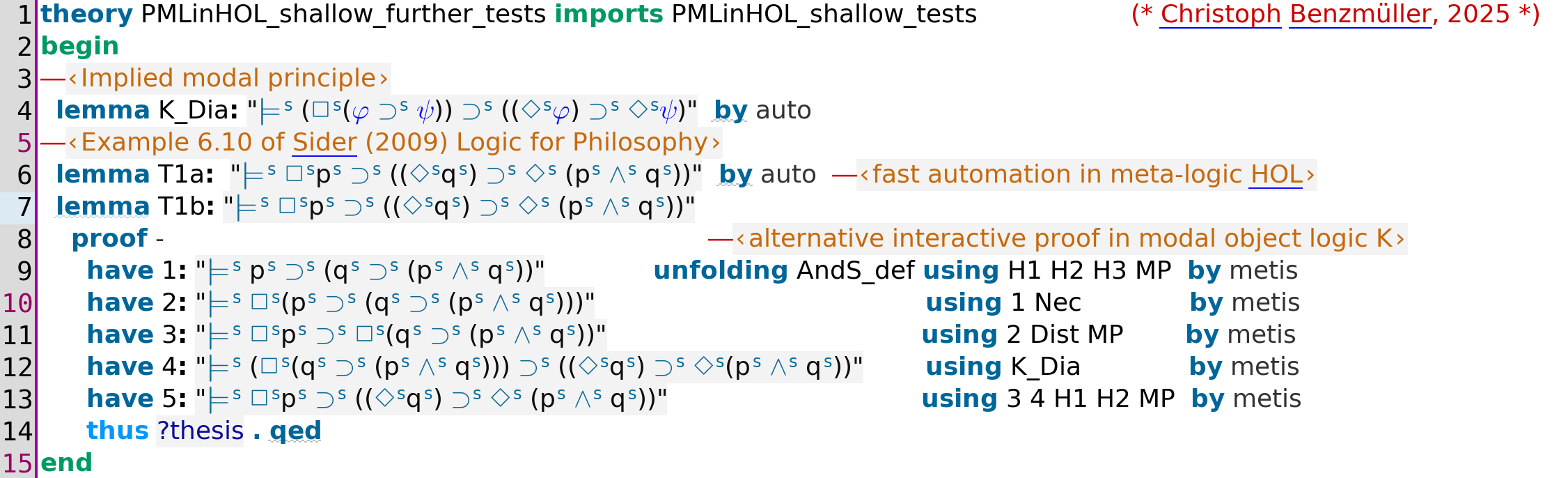}}
   \vskip.5em
  \colorbox{gray!30}{\includegraphics[width=.97\textwidth]{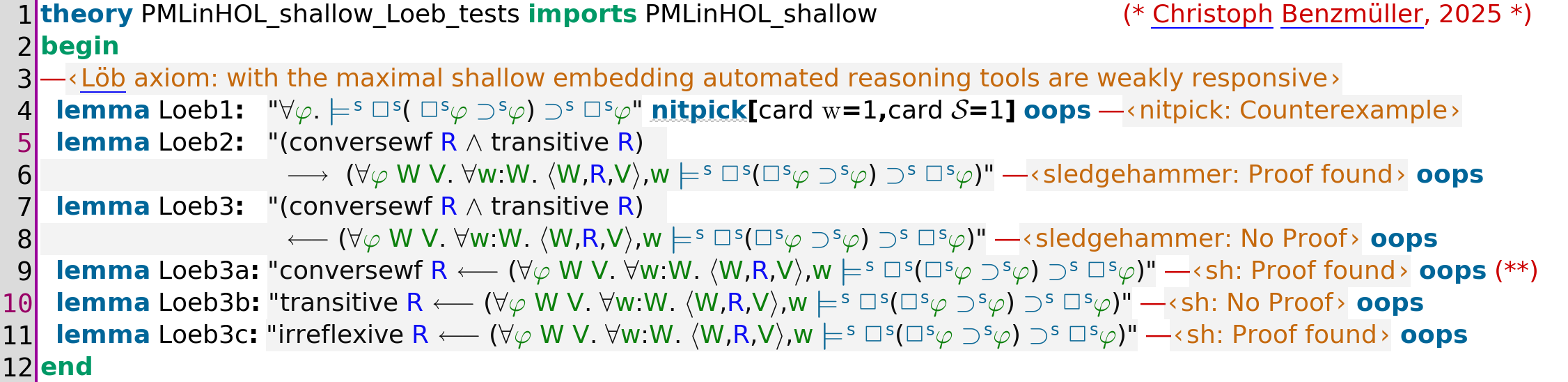}}
  \caption{Experiments with the maximal shallow embedding of PML in HOL. \label{fig:PMLinHOL_shallow}}
\end{figure}

\begin{figure}[htp!]
  \centering
  \colorbox{gray!30}{\includegraphics[width=.97\textwidth]{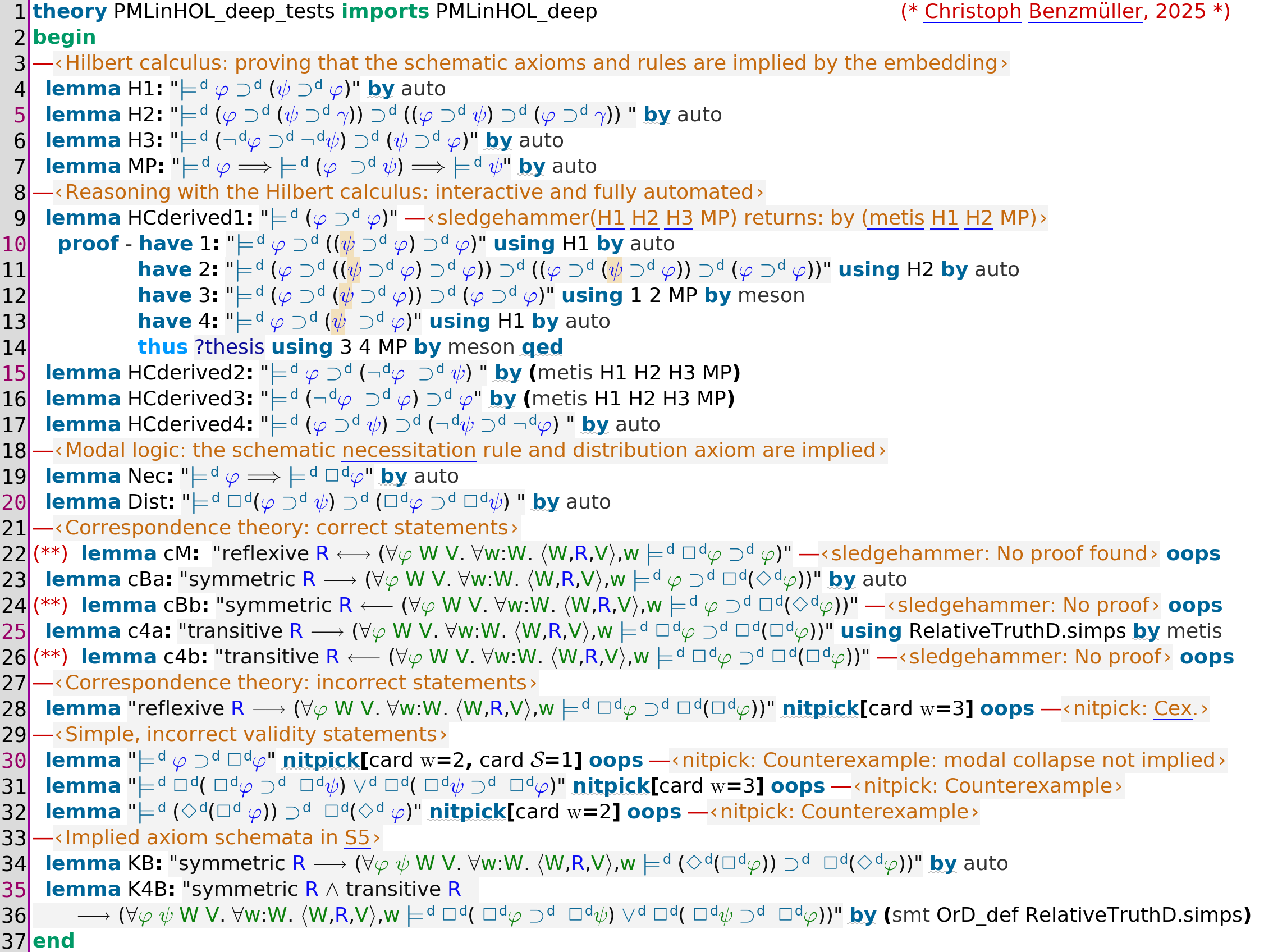}}
  \vskip.5em
  \colorbox{gray!30}{\includegraphics[width=.97\textwidth]{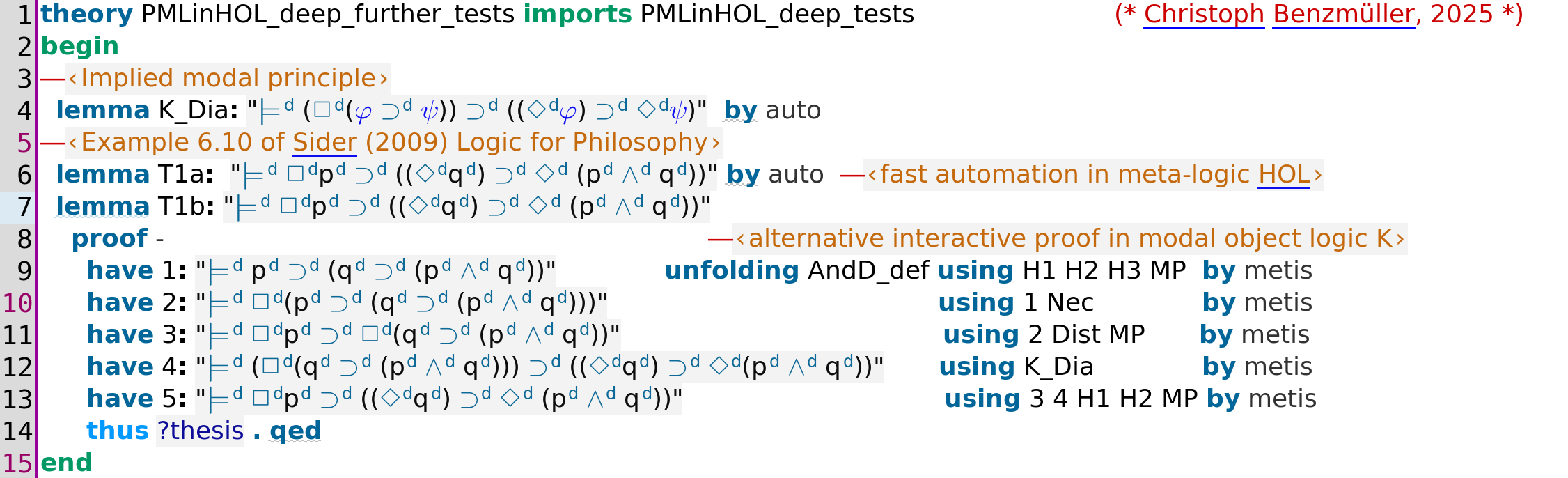}}
   \vskip.5em
  \colorbox{gray!30}{\includegraphics[width=.97\textwidth]{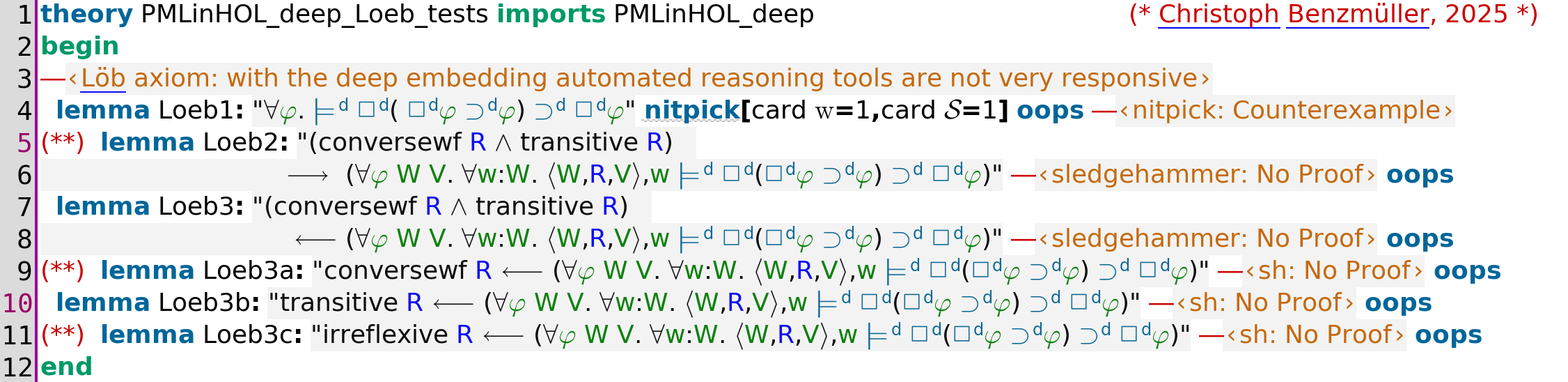}}
  \caption{Experiments with the deep embedding of PML in HOL. \label{fig:PMLinHOL_deep}}
\end{figure}

Experiments with the maximal shallow embedding 
and the deep embedding of PML in HOL 
are presented in Figs.~\ref{fig:PMLinHOL_shallow} and ~\ref{fig:PMLinHOL_deep} respectively. Automation results that divert from those reported for the minimal shallow embedding are highlighted by $\textcolor{red}{(**)}$.
\newpage

\newpage

\end{appendix}
\end{document}